\documentclass[11pt]{article}
\usepackage[utf8]{inputenc}
\usepackage{amsmath,amsfonts,amssymb}
\usepackage{amsthm}
\usepackage{latexsym}
\usepackage[noadjust]{cite}
\usepackage{graphicx}
\usepackage{xcolor}
\usepackage{dirtytalk}
\usepackage{mathtools}
\usepackage[margin=1in]{geometry}
\usepackage{thm-restate}
\usepackage{enumitem}
\usepackage[linesnumbered,ruled]{algorithm2e}
\usepackage[colorlinks=true, allcolors=blue]{hyperref}
\usepackage[nameinlink,capitalise]{cleveref}
\hypersetup{
    citecolor={violet}
}

\newtheorem{theorem}{Theorem}[section]

\newtheorem{lemma}[theorem]{Lemma}

\newtheorem{claim}[theorem]{Claim}
\theoremstyle{definition}
\newtheorem{definition}{Definition}[section]
\newtheorem{remark}[theorem]{Remark}

\newcommand{\F}{\mathbb{F}}
\newcommand{\N}{\mathbb{N}}

\newcommand{\R}{\mathbb{R}}

\newcommand{\zo}{\{0, 1\}}

\newcommand{\eps}{\epsilon}

\newcommand{\wt}{\mathrm{wt}}

\newcommand{\Supp}{\mathrm{Supp}}

\title{A Theory of Spectral CSP Sparsification}
\date{\today}
\author{Sanjeev Khanna\thanks{School of Engineering and Applied Sciences, University of Pennsylvania, Philadelphia, PA. Supported in part by NSF award CCF-2402284 and AFOSR award FA9550-25-1-0107. Email: {\tt sanjeev@cis.upenn.edu}. } \and Aaron (Louie) Putterman\thanks{School of Engineering and Applied Sciences, Harvard University, Cambridge, Massachusetts, USA. Supported in part by the Simons Investigator Awards of Madhu Sudan and Salil Vadhan, NSF Award CCF 2152413 and AFOSR award FA9550-25-1-0112. Email: \texttt{aputterman@g.harvard.edu}.} \and Madhu Sudan\thanks{School of Engineering and Applied Sciences, Harvard University, Cambridge, Massachusetts, USA. Supported in part by a Simons Investigator Award, NSF Award CCF 2152413 and AFOSR award FA9550-25-1-0112. Email: \texttt{madhu@cs.harvard.edu}.}}

\begin{document}

\maketitle

\begin{abstract}
We initiate the study of spectral sparsification for instances of Constraint Satisfaction Problems (CSPs). In particular, we introduce a notion of the \emph{spectral energy} of a fractional assignment for a Boolean CSP instance, and define a \emph{spectral sparsifier} as a weighted subset of constraints that approximately preserves this energy for all fractional assignments. Our definition not only strengthens the combinatorial notion of a CSP sparsifier but also extends well-studied concepts such as spectral sparsifiers for graphs and hypergraphs.

Recent work by Khanna, Putterman, and Sudan [SODA 2024] demonstrated near-linear sized \emph{combinatorial sparsifiers} for a broad class of CSPs, which they term \emph{field-affine CSPs}. Our main result is a polynomial-time algorithm that constructs a spectral CSP sparsifier of near-quadratic size for all field-affine CSPs. This class of CSPs includes graph (and hypergraph) cuts, XORs, and more generally, any predicate which can be written as $P(x_1, \dots x_r) = \mathbf{1}[\sum a_i x_i \neq b \mod p]$.

   Based on our notion of the spectral energy of a fractional assignment, we also define an analog of the second eigenvalue of a CSP instance. We then show an extension of Cheeger's inequality for all even-arity XOR CSPs, showing that this second eigenvalue loosely captures the ``expansion'' of the underlying CSP. This extension specializes to the case of Cheeger's inequality when all constraints are even XORs and thus gives a new generalization of this powerful inequality which converts the combinatorial notion of expansion to an analytic property.

   Perhaps the most important effect of spectral sparsification is that it has led to certifiable sparsifiers for graphs and hypergraphs. This aspect remains open in our case even for XOR CSPs since the eigenvalues we describe in our Cheeger inequality are not known to be efficiently computable. Computing this efficiently, and/or finding other ways to certifiably sparsify CSPs are open questions emerging from our work. Another important open question is determining which classes of CSPs have near-linear size \emph{spectral} sparsifiers.
\end{abstract}
\pagenumbering{gobble}

\pagebreak

\tableofcontents

\pagebreak

\pagenumbering{arabic}

\section{Introduction}

Spectral tools have played a powerful role in the analysis of graphs and hypergraphs and led to many celebrated successes including the use of Cheeger's inequality to approximate computationally intractable notions like expansion \cite{AM85, Alo86} and the use of spectral methods to sparsify graphs and hypergraphs \cite{BSS09, SY19}. Recent years have seen the use of CSPs to add greater richness to the study of hypergraphs by considering generalized notion of when an edge (or hyperedge) is deemed to be cut by a (multi-)partition of the vertex set. In this work we initiate the study of spectral methods for CSP analysis, by introducing a concrete notion of the spectral energy of a (fractional) assignment to variables, and giving analogs to Cheeger's inequality and spectral sparsification. We describe our work in more detail below after reviewing some of the classical works on spectral analysis, notably on sparsification.

\subsection{Background}

Given a graph $G = (V, E)$, there are many important combinatorial properties of the graph that are hard to explicitly calculate. Perhaps most notably is the notion of \emph{conductance}: given a graph $G = (V, E)$, we can define its conductance as 
\[
\phi_G = \min_{S \subset V} \frac{|\delta S|}{\min(\mathrm{Vol}(S),\mathrm{Vol}(V - S))}.
\]
Here, $\mathrm{Vol}(S)$ refers to the sum of degrees of vertices in $S$, and $\delta S$ refers to the edges crossing from $S$ to $\bar{S}$. In a celebrated result \cite{AM85, Alo86}, it was shown that $\phi_G$ can actually be well-approximated by the second eigenvalue of the normalized Laplacian of the graph: 
\[
\frac{\lambda_2}{2} \leq \phi_G \leq \sqrt{2 \lambda _2}.
\]
The normalized Laplacian is simply $\mathcal{L} = W^{-1/2} (W-A)W^{-1/2}$, where $W$ is the diagonal matrix of vertex degrees, and $A$ is the (weighted) adjacency matrix of the graph. Subsequent to this discovery, spectral theory has emerged as a central topic in graph theory with wide-reaching implications and connections (see \cite{Spi07} for further discussion). In particular, one line of research emanating from spectral theory has been \emph{spectral sparsification}, which is the key focus of this paper.

\subsubsection{(Spectral) Graph Sparsification}

Graph sparsification has emerged as a key topic in algorithmic graph theory. First proposed in the works of Karger \cite{Kar93} and Bencz\`ur and Karger \cite{BK96}, graph sparsification takes as input a graph $G = (V, E)$ and a parameter $\eps \in (0,1)$, and returns a reweighted sub-graph $G'$ such that the size of every cut in $G$ is simultaneously preserved to a $(1 \pm \eps)$ multiplicative factor in $G'$. Although the algorithm presented in \cite{BK96} is nearly-optimal for cut sparsification, these works spurred a flurry of research in sparsification in many different directions: for instance, the design of \emph{deterministic} algorithms for graph sparsification \cite{BSS09}, generalizing notions of sparsification to hypergraphs \cite{KK15, CKN20, KKTY21b} and to norms \cite{JLLS23}, to CSPs \cite{KK15, KPS24}, and to arbitrary sets of vectors \cite{BrakensiekG}. Along the way, one of the core advances in the study of sparsification has been the notion of a \emph{spectral} sparsifier \cite{ST11}. Here, instead of only preserving the \emph{cut-sizes} of a graph, the sparsifier instead preserves the \emph{energy} of the graph, which is defined for a graph $G = (V, E)$, and a vector $x \in [0,1]^V$ as $Q_G(x) = \sum_{e = (u,v) \in E} w_e \cdot (x_u - x_v)^2$. Specifically, a spectral sparsifier is a re-weighted subgraph $G'$ such that \emph{for every} $x \in [0,1]^V$, it is the case that $Q_{G'}(x) \in (1 \pm \eps) Q_G(x)$. This energy can be written as a quadratic form involving the graph's (un-normalized) Laplacian, which is denoted by $L_G$, and satisfies $(L_G)_{i,j} = \begin{cases}
    \mathrm{deg}(i) \quad \text{if } i = j \\
    -w_{i,j} \quad \text{ else}
\end{cases}.$

In graphs, the benefits from the use of spectral sparsification are several-fold:
\begin{enumerate}
\item Spectral sparsification is a stronger condition than cut sparsification, as it preserves the eigenvalues of the graph's Laplacian in addition to the cut sizes.
    \item The energy of a graph allows for an interpretation of the graph as an electrical network, which allows for well-motivated sampling schemes for designing sparsifiers.
    \item The energy of the graph has a convenient form as the quadratic form of the Laplacian of the graph. This means that spectral sparsifiers are \emph{efficiently certifiable} (in the sense that one can verify in polynomial time whether a graph $G'$ is a $(1 \pm \eps)$ spectral-sparsifier of a graph $G$), a key fact that is used when designing deterministic sparsification algorithms \cite{BSS09}.
    \item This property of efficient (and deterministic) certification has also contributed to \emph{linear-size} spectral sparsifiers of graphs \cite{BSS09}, shaving off the $\log(n)$ factor that is inherent to the sampling-based approaches that create cut-sparsifiers.
\end{enumerate}

\subsubsection{(Spectral) Hypergraph Sparsification}
Given the wide-reaching benefits of spectral sparsification, as research on graph sparsification pivoted to \emph{hypergraph} sparsification, spectral notions of sparsification in hypergraphs were quick to be proposed. In a hypergraph $H = (V, E)$, each hyperedge $e \in E$ is an arbitrary subset of vertices. A cut in the hypergraph $H$ is given by a subset $S \subseteq V$ of the vertices, and we say that a hyperedge $e$ is cut by $S$ if $S \cap e \neq \emptyset$ and $\bar{S} \cap e \neq \emptyset$. A $(1 \pm \eps)$ cut-sparsifier of a hypergraph is then just a re-weighted subset of hyperedges which simultaneously preserves the weight of every cut to a $(1 \pm \eps)$ factor. Analogous to the graph case, spectral sparsifiers of hypergraphs instead operate with an \emph{energy} formulation. Given a hypergraph $H = (V, E)$ and a vector of potentials $x \in [0,1]^V$, the energy of the hypergraph is given by $Q_H(x) = \sum_{e \in E} w_e \cdot \max_{u,v \in e} (x_u - x_v)^2$, and, as with graphs, a spectral sparsifier is simply a re-weighted subset of hyperedges which preserves this energy to a $(1 \pm \eps)$ factor \emph{simultaneously} for every $x$.

A line of works by Louis and Makarychev \cite{LM14}, Louis \cite{Lou15}, and Chan, Louis, Tang, and Zhang \cite{CLTZ18} showed that this energy definition in hypergraphs enjoys many of the same properties as the energy definition in graphs: in particular, that it can be viewed as a random walk operator on the hypergraph, that it generalizes the cuts of a hypergraph, and that it admits Cheeger's inequalities in a manner similar to graphs. Because of these connections, a long line of works simultaneously studied the ability to \emph{cut-sparsify} (\cite{KK15, CKN20}) and \emph{spectrally-sparsify} (\cite{SY19, BST19, KKTY21a, KKTY21b, JLS22, Lee23}) hypergraphs.

\subsubsection{CSP Sparsification} 

However, subsequent to this unified effort to understand hypergraph sparsification in the cut and spectral settings, the directions of focus from the sparsification community have largely been separate; on the one hand, combinatorial generalizations of cut-sparsification have been studied for more general objects like linear (and non-linear) codes as well as CSPs \cite{KPS24, KPS24c, BrakensiekG}, while on the other hand, continuous generalizations of spectral sparsification have been studied in the context of semi-norms and generalized linear models \cite{JLLS23, JLLS24}. The techniques used by these respective research efforts are likewise separate, with combinatorial sparsifiers driven mainly by progress on counting bounds, sampling, and union bounds, while continuous sparsifiers have relied on a deeper understanding of chaining. 

In this work, we aim to close this gap between recent work on combinatorial and continuous notions of sparsification by introducing the model of \emph{spectral CSP sparsification}. This framework simultaneously generalizes cut and spectral sparsification of graphs, cut and spectral sparsification of hypergraphs, and the (combinatorial) sparsification of codes and CSPs. We then show that even under this more general definition, there still exist sparsifiers of small sizes, and moreover, these sparsifiers can even be computed efficiently. We summarize our contributions more explicitly in the following subsection.

\subsection{Our Contributions}

To start, we formally define CSP sparsification, as it will be a key building block in our discussion of spectral CSPs.

\begin{definition}
    A CSP $C$ on $n$ variables and $m$ constraints is given by a predicate $P: \zo^r \rightarrow \zo$, along with $m$ ordered subsets of variables, denoted $S_1, \dots S_m$ where each $S_j \in [n]^r$. The $j$th constraint is then denoted by $P(x_{S_j})$, where the arguments in $P$ are understood to be $\{x_{\ell}: \ell \in S_j\}$. The CSP $C$ can also be accompanied by weights $w_c$ for each constraint $c \in C$. Often, for a constraint $c \in C$, we will use $c(x)$ to denote the evaluation of the constraint $c$ on assignment $x$. Here it is understood that $c$ may only be acting on a \emph{subset} of the variables in the assignment $x$.
\end{definition}

\begin{definition}
    Let $C$ be a CSP on $n$ variables and $m$ constraints. For an assignment $x \in \zo^n$, the value of the assignment is
    \[
    |C(x)| = \sum_{j = 1}^m w_j \cdot P(x_{S_j}).
    \]
    In words, this is simply the weight of all satisfied constraints. 
\end{definition}

\begin{definition}
    For a CSP $C$ on $n$ variables and $m$ constraints, a $(1 \pm \eps)$ sparsifier $\hat{C}$ of $C$ is a re-weighted CSP $\hat{C}$, with weights $\hat{w}_j: j \in [m]$ such that $\forall x \in \zo^n$:
    \[
    |\hat{C}(x)| \in (1 \pm \eps) \cdot |C(x)|.
    \]
    The sparsity of $\hat{C}$ is then given by $|\hat{w}|_0$.
\end{definition}

As remarked in many previous works, when the predicate $P: \zo^2 \rightarrow \zo$ is the $2$-XOR function, then CSP sparsification captures the notion of cut-sparsification in ordinary graphs. This is because a constraint $x_i \oplus x_j$ simulates an edge $(i,j)$ and will evaluate to $1$ if and only if $x_i \neq x_j$ (equivalently, if $i,j$ are on different sides of the cut - see \cite{KPS24} for further discussion). Likewise, when the predicate $P: \zo^r \rightarrow \zo$ is the \emph{not-all-equal} function (i.e., evaluates to $0$ on $0^r$ and $1^r$), then this exactly captures the notion of a hyperedge of arity $r$ being cut. Now, recall that in graphs, the corresponding energy of an edge $(i,j)$ is $(x_i - x_j)^2$. Spectral sparsification of graphs thereby captures cut sparsification as when $x \in \zo^n$, then $(x_i - x_j)^2 = x_i \oplus x_j$. For arbitrary CSPs, we capture this behavior as follows:

\begin{definition}
    For a vector $x \in [0,1]^n$, and a value $\theta \in [0,1]$, the (deterministic) rounding of $x$ with respect to $\theta$ is the vector $x^{(\theta)}$ such that
    \[
    x^{(\theta)}_i = \mathbf{1}[x_i \geq \theta].
    \]
\end{definition}

This leads to a straightforward definition of the energy of a CSP:

\begin{definition}
    For a CSP $C$ on $n$ variables and $m$ constraints and a vector $x \in [0,1]^n$, the energy of $C$ is defined as 
    \[
    Q_C(x) = \sum_{c \in C} w_c \cdot \left(\Pr_{\theta \in [0,1]}[c(x^{(\theta)}) = 1] \right)^2.
    \]
\end{definition}

Our notion of spectral sparsification is then a natural continuation of this idea:

\begin{definition}
    For a CSP $C$ on $n$ variables and $m$ constraints, we say that a re-weighted sub-CSP $\hat{C}$ is a $(1 \pm \eps)$ spectral-sparsifier of $C$ if $\forall x \in [0,1]^n$
    \[
    Q_{\hat{C}}(x) \in (1 \pm \eps)Q_{C}(x).
    \]
\end{definition}

For instance, let us consider an assignment of potentials $x \in [0,1]^n$, and a constraint $c(x) = x_i \oplus x_j$. Now, our goal is to understand what the expression $\Pr_{\theta \in [0,1]}[c(x^{(\theta)}) = 1]$ looks like; indeed, the only case where the constraint evaluates to $1$ is when $\theta$ falls in the interval $[x_i, x_j]$. If $\theta$ is smaller than $\min\{ x_i, x_j \}$, then both become $1$ in the rounded vector and the XOR becomes $0$, and if $\theta$ is larger than $\max\{ x_i, x_j \}$, then they both become $0$. Thus, the only way for the XOR constraint to become $1$ is if $\theta$ is larger than exactly one of $x_i, x_j$: in this case one of the variables gets rounded to $1$, and the other gets rounded to $0$. Thus, $\Pr_{\theta \in [0,1]}[c(x^{(\theta)}) = 1] = |x_i - x_j|$, and when we consider this expression squared, we recover $(x_i - x_j)^2$, which exactly mirrors the energy for ordinary graphs. A similar analysis shows that when we consider $r$-NAE constraints, we recover exactly the same energy expression as in hypergraphs of arity $r$. 

Naturally, in the context of sparsification, the next question to ask is whether this more general notion of CSP sparsification still allows for sparsifiers of small sizes. Our first result shows that for a broad class of CSPs, this is indeed the case. Specifically, we build upon the work of \cite{KPS24}, and consider \emph{field-affine} CSPs; namely CSPs using a predicate $P$ where $P(x_1, \dots x_r)  = \mathbf{1}[\sum_{i} a_i x_i \neq b_i \mod q]$ for a prime $q$. For these CSPs, we show the following:

\begin{theorem}\label{thm:generalCSPintro}
	Let $C$ be any CSP on $n$ variables and $m$ constraints using a predicate $P: \zo^r \rightarrow \zo$ such that $P(y) = 1 \iff \mathbf{1}[a_i y_i \neq b \mod p]$, for some prime $p$. Then, there is a randomized, polynomial time algorithm which with high probability computes a $(1 \pm \eps)$ spectral-sparsifier of $C$ that retains only $\widetilde{O}(n^2 \log^2(p) / \eps^2)$ re-weighted constraints.
\end{theorem}

Note that this theorem encapsulates a large variety of predicates, including arbitrary arity XORs, graph and hypergraph cut functions, and hedge-graph cut functions \cite{KPS24}. Additionally, while there has been a large research effort for sparsifying continuous functions \cite{JLLS23, JLLS24}, many classes of CSPs do not fit into the established frameworks. For instance, if we consider a $4$-XOR constraint $c(x_1, x_2, x_3, x_4) = x_1 \oplus x_2 \oplus x_3 \oplus x_4$, then we can immediately observe that for the assignment $x_1 = x_2 = 0, x_3 = x_4 = 1/2$, $\Pr_{\theta \in [0,1]}[c(x^{(\theta)}) = 1] = 0$, as there is no choice of $\theta$ for which an odd number of the $x_i$'s round to $1$. The same holds true when we consider the assignment $x_1 = x_3 = 0, x_2 = x_4 = 1/2$. However, when we \emph{add} these assignments together, yielding $x_1 = 0, x_2 = x_3 = 1/2, x_4 = 1$, then in fact $\Pr_{\theta \in [0,1]}[c(x^{(\theta)}) = 1] = 1$, as there is always an odd number of $x_i$'s that round to $1$, for any choice of $\theta$. All together, this means that the XOR functions \emph{strongly disobey} the triangle inequality, which is one of the key properties that results on sparsifying sums of continuous functions rely on \cite{JLLS23, JLLS24}. Nevertheless, \cref{thm:generalCSPintro} shows that these functions still admit sparsifiers of small size, suggesting that there may be other avenues towards showing the sparsifiability of sums of continuous functions.

Finally, we show that despite the generality of our definition of the spectrum of a CSP, our definitions still admit many of the same properties enjoyed by the spectra of graphs and hypergraphs. In particular, we prove that a type of \emph{Cheeger inequality} holds, relating eigenvalues of the CSP to a combinatorial notion of the \emph{expansion} of the CSP. Our Cheeger inequality is defined with respect to the eigenvalues of a discrepancy ratio (defined analogously to \cite{CLTZ18}), and we use $\gamma_2$ to denote the second smallest eigenvalue. Likewise, we define the expansion in an intuitive way to generalize graphs and hypergraphs: the expansion is typically measured as a ratio of the number of edges \emph{leaving} a set $S$ of vertices divided by the number of edges \emph{inside} this same set. Under a CSP perspective, a constraint $c$ is considered to be \emph{leaving} a set $S$ if the constraint $c$ evaluates to $1$ when we apply the assignment $\mathbf{1}_S$. For example, in graphs, each edge $(u,v)$ can be modeled by an XOR of the variables $x_u$ and $x_v$. A simple check shows that the edge $(u,v)$ is only crossing between $S$ and $\bar{S}$ if $(\mathbf{1}_S)_u \oplus (\mathbf{1}_S)_v = 1$. We defer formal definitions of these notions to Section~\cref{sec:cheeger}. Assuming we denote the expansion of a CSP $C$ by $\Phi_C$, we establish the following theorem:

\begin{theorem}
    Given any XOR CSP $C$ where each constraint is of even size and maximum arity $\ell$,
    \[
    \frac{\gamma_2}{2} \leq \Phi_C \leq \left (2 \sqrt{\ell/2} + 1 \right) \sqrt{\gamma_2},
    \]
    where $\gamma_2$ is the eigenvalue of the XOR-CSP Laplacian defined in \cref{def:eigenvalue}. 
\end{theorem}

\subsection{Technical Overview}

In this subsection, we summarize the main ideas that go into our proof of the ability to spectrally sparsify certain classes of CSPs.

\subsubsection{Writing as a Quadratic Form}

The starting point for our approach is a technique used in the work of Soma and Yoshida \cite{SY19}. To convey the main idea, we will consider the simplified setting of spectrally sparsifying XOR constraints of arity $4$. That is, we consider the predicate $P(y_1, y_2, y_3, y_4) = y_1 \oplus y_2 \oplus y_3 \oplus y_4$, and each constraint $c_i(x) = P(x_{i_1}, x_{i_2}, x_{i_3}, x_{i_4})$. We denote the entire CSP by $C$, which consists of constraints $c_1, \dots c_m$, and we denote the set of variables by $x_1, \dots x_n$. We will often index the set $[m]$ by a constraint $c \in C$ (as there are $m$ different constraints). Note that we choose arity $4$ XORs instead of $3$ so that the all $1$'s assignment is unsatisfying (and thus we can shift any assignment by the all $1$'s vector WLOG). We later show that arity $3$ XORs can be simulated by arity $4$ XORs.

Recall then that our goal is to sparsify the set of constraints while still preserving the energy of the CSP to a $(1 \pm \eps)$ factor, where the energy is exactly 
\[
Q_C(x) = \sum_{c \in C} w_c \cdot \left(\Pr_{\theta \in [0,1]}[c(x^{(\theta)}) = 1] \right)^2.
\]
Now, when we focus on a single constraint, we can actually interpret this energy definition concretely. Since a single constraint is a $4$-XOR over variables $x_{i_1}, \dots x_{i_4}$, then with respect to a choice of the rounding parameter $\theta$, the constraint evaluates to $1$ if and only if an odd number of the variables $x_{i_1}, \dots x_{i_4}$ are larger than the rounding threshold $\theta$. Thus, if we sort $x_{i_1}, \dots x_{i_4}$ according to their value (say the sorted order is $x_{i_{o(1)}} \leq x_{i_{o(2)}} \leq x_{i_{o(3)}} \leq x_{i_{o(4)}}$) the constraint evaluates to $1$ if and only if $\theta \in (x_{i_{o(1)}}, x_{i_{o(2)}})$ or $(x_{i_{o(3)}}, x_{i_{o(4)}})$. Because $\theta$ is chosen uniformly at random from $[0,1]$, we can directly calculate this probability to be
\[
\Pr_{\theta \in [0,1]}[c(x^{(\theta)}) = 1] = x_{i_{o(4)}} -x_{i_{o(3)}} + x_{i_{o(2)}} - x_{i_{o(1)}}.
\]
Ultimately, the energy is the \emph{square} of this expression, and so it can be written as 
\[
\left ( \Pr_{\theta \in [0,1]}[c(x^{(\theta)}) = 1] \right )^2 = \left ( x_{i_{o(4)}} -x_{i_{o(3)}} + x_{i_{o(2)}} - x_{i_{o(1)}} \right )^2.
\]

Here is where we use a simple observation: for a \emph{fixed} ordering of the variables $\pi$ (i.e., such that $x_{\pi(1)} \leq x_{\pi(2)} \leq \dots \leq x_{\pi(n)}$), the energy expression for each constraint evaluates to a fixed quadratic form. We denote this set of vectors that satisfies the ordering $\pi$ by $[0,1]^{\pi}$, and call a vector $x$ satisfying $x_{\pi(1)} \leq x_{\pi(2)} \leq \dots \leq x_{\pi(n)}$ $\pi$-consistent.
Said another way, once we fix the ordering of the variables, then $o(1), o(2), o(3), o(4)$ in the above expression would also all be fixed. For such a permutation $\pi$, we call this resulting quadratic form the \emph{induced quadratic form by $\pi$.} For example, if we take the ordering $\pi$ to be the identity permutation (i.e., $x_1 \leq x_2 \leq \dots \leq x_n$), and we suppose that $i_1 < i_2 < i_3 < i_4$, then the formula for the energy of the expression would be exactly $(x_{i_4} - x_{i_3}  + x_{i_2} - x_{i_1})^2$. Note that our choice of the words \emph{quadratic form} is intentional, as once the energy expression is a fixed sum of these quadratic terms, we can express the energy as $Q_{C}(x) = x^T (B^{\pi})^T W_C B^{\pi} x$, where $x \in [0,1]^{\pi}$, and $B^{\pi} \in \{-1,0,1\}^{m \times n}$ is the matrix such that $((B^{\pi}) x)_i = \Pr_{\theta \in [0,1]}[c_i(x^{(\theta)}) = 1]$, and $W_C$ is simply the diagonal matrix whose $i,i$th entry is the weight of the $i$th constraint in $C$.

\subsubsection{Sufficient Conditions for Preserving Quadratic Forms}
Now, as observed in \cite{SY19}, in order to sparsify the CSP $C$, it suffices to choose a re-weighted subset of indices $\hat{S} \subseteq [m]$, such that \emph{simultaneously for every ordering} $\pi$, the induced quadratic form by $\pi$ has its energy preserved to a $(1 \pm \eps)$ factor. Because of our above simplifications, we can re-write this another way: we want to find a new, sparser set of weights on our constraints (which we denote by $\widehat{C}$), such that \emph{for every} ordering $\pi$, and \emph{for every} vector $x \in [0,1]^{\pi}$, it is the case that 
	\[
	Q_{\hat{C}}(x) = x^T (B^{\pi})^T W_{\hat{C}} B^{\pi} x \in (1 \pm \eps) x^T (B^{\pi})^T W_{C} B^{\pi}  x = (1 \pm \eps ) Q_{C}(x).
	\]	
	We can re-write this condition as saying that 
	\[
	 x^T \left ( (B^{\pi})^T W_{\hat{C}} B^{\pi} - (1 - \eps)(B^{\pi})^T W_{C} B^{\pi}  \right ) x \geq 0,
	\]
	and 
	\[
		 x^T \left ( (1 + \eps)(B^{\pi})^T W_{C} B^{\pi}  - (B^{\pi})^T W_{\hat{C}} B^{\pi}  \right ) x \geq 0.
	\]
If this condition holds, then we will have indeed created a $(1 \pm \eps)$ spectral sparsifier. Recall that we want to use \emph{as few} re-weighted constraints as possible, and so we want this new set of re-weighted constraints $\hat{C}$ to be as sparse as possible. 

Next, observe that the expression above is a type of \emph{positive definiteness} condition, where we focus our attention on vectors in $[0,1]^{\pi}$, and want to ensure that the quadratic form of some matrix is non-negative. In particular, the work of \cite{SY19} studied similar matrices, and gave an exact characterization for when such a claim holds:

\begin{lemma}[Lemma 3.3 in \cite{SY19}]\label{lem:SYintro}
	Let $A \in \R^{n \times n}$ be a symmetric matrix, such that $A \cdot \mathbf{1} = 0$, and let $\pi$ be a permutation on $[n]$. Then, $x^TAx \geq 0$ for all $x \in [0,1]^{\pi}$ if and only if $A$ is of the form $P_{\pi}^T J^T K J P_{\pi}$ for a matrix $K \in \R^{(n-1) \times (n-1)}$ such that $\forall y \in \R_+^n, y^T K y \geq 0$, and where  $P_{\pi} \in \R^{n \times n}$ is the matrix such that $P_{\pi}(i,j) = 1$ if $j = \pi_i$, and otherwise is $0$ and $J \in \R^{(n-1) \times n}$, where $J(i,i) = 1$, $J(i, i+1) = -1$, and otherwise is $0$.
\end{lemma}

This lemma gives us a roadmap for how to proceed. Recall that there are two matrices that we wish to show satisfy the positivity condition, namely $\left ( (B^{\pi})^T W_{\hat{C}} B^{\pi} - (1 - \eps)(B^{\pi})^T W_{C} B^{\pi}  \right )$ and $ \left ( (1 + \eps)(B^{\pi})^T W_{C} B^{\pi}  - (B^{\pi})^T W_{\hat{C}} B^{\pi}  \right )$,  and these will each be the matrix $A$ in the above lemma (though to simplify discussion, we will simply focus on the first, as the analysis will be identical). Our goal is to show that we can re-write the matrix as
\[
\left ( (B^{\pi})^T W_{\hat{C}} B^{\pi} - (1 - \eps)(B^{\pi})^T W_{C} B^{\pi}  \right ) = P_{\pi}^T J^T K J P_{\pi},
\]
where the matrix $K$ is \emph{co-positive} (meaning for all positive vectors, the quadratic form is non-negative). To do this, we will ultimately define a new, specific matrix $B_{\mathrm{cross}}^{\pi}$ with the intention that $B^{\pi} = B_{\mathrm{cross}}^{\pi} J P_{\pi}$. While \cite{SY19} also define a crossing matrix, ours is necessarily different as it models a different structure. It is here where our analysis begins to diverge from theirs. 

\subsubsection{Building the Crossing Matrix}

However, before defining this crossing matrix $B_{\mathrm{cross}}^{\pi}$, we first require some definitions. First is the notion of an \emph{active region}, which is essentially the intervals where the contribution to the energy comes from (i.e., the choices of $\theta$ for which a constraint is satisfied):

\begin{definition}
	For a fixed ordering $\pi$ on $[n]$, and a 4-XOR constraint $c$ operating on elements $x_{i_1} \leq x_{i_2} \leq x_{i_3} \leq x_{i_4}$, we say the active regions are 
	\[
	[x_{i_1}, x_{i_2}] \cup [x_{i_3}, x_{i_4}]
	\]
\end{definition}

For an index $i \in [n-1]$, let us consider the bisector between $x_{\pi_i}$ and $x_{\pi_{i+1}}$ (i.e., between the $i$th and $i+1$st smallest elements in the ordering $\pi$). This bisector takes on value $(x_{\pi_i} + x_{\pi_{i+1}}) / 2$. We use this to define \emph{crossing indices}:

\begin{definition}
	For $i \in [n-1]$, we say that $i$ is a crossing index with respect to the ordering $\pi$ and constraint $c$ if the bisector $(x_{\pi_i} + x_{\pi_{i+1}}) / 2$ is in an active region of the constraint $c$ with ordering $\pi$.
\end{definition}

Likewise, we generalize this definition to a pair of indices $i,j \in [n-1]$:

\begin{definition}
	We say that pairs of indices $i,j \in \binom{[n-1]}{2}$ are crossing indices with respect to the constraint $c$, ordering $\pi$ if both $i, j$ individually are crossing indices with $c, \pi$.
\end{definition}

For instance, let us consider what happens when the permutation $\pi = \mathrm{Id}_n$, and suppose we are looking at a constraint $c_i(x)$ which is the XOR of $x_{2}, x_{3}, x_{6}, x_{9}$. Then the crossing indices are exactly $2, 6, 7$ and $8$. More intuitively, the crossing indices correspond to the intervals $[x_{i}, x_{i+1}]$ which contribute to the overall energy (i.e., for which choices of $\theta$ does the constraint evaluate to $1$). For instance, because $[x_6, x_9]$ is an active region the energy contributed from these points is $x_9 - x_6$ and so every sub-interval $[x_6, x_7], [x_7, x_8], [x_8, x_9]$ contributes to the energy. This is because if $\theta \in [x_7, x_8]$ for instance, then in the rounding $x_2 = x_3 = x_6 = 1$, and $x_9 = 0$, and hence the constraint is satisfied. 

With these definitions established, we define the crossing matrix $B_{\mathrm{cross}}^{\pi} \in \R^{m \times n-1}$ such that $B_{\mathrm{cross}}^{\pi}(c, i) = 1$ if and only if $i$ is a crossing index for the constraint $c$ under permutation $\pi$ (and otherwise, the $c, i$th entry is $0$). In particular, this specific definition allows us to show that
\[
B^{\pi} = B_{\mathrm{cross}}^{\pi} J P_{\pi}.
\]
We do not discuss this equality exactly here, as it requires an extensive case analysis; we defer this to the technical sections below. 

However, using this, we can re-write our original expression:
\[
\left ( (B^{\pi})^T W_{\hat{C}} B^{\pi} - (1 - \eps)(B^{\pi})^T W_{C} B^{\pi}  \right ) = P_{\pi}^T J^T ( (B_{\mathrm{cross}}^{\pi})^T W_{\hat{C}} B_{\mathrm{cross}}^{\pi} - (1 - \eps)(B_{\mathrm{cross}}^{\pi})^T W_{C} B_{\mathrm{cross}}^{\pi}) J P_{\pi}.
\]
Thus, if we revisit \cref{lem:SYintro}, the matrix $K$ is exactly this interior portion of the expression $K = ( (B_{\mathrm{cross}}^{\pi})^T W_{\hat{C}} B_{\mathrm{cross}}^{\pi} - (1 - \eps)(B_{\mathrm{cross}}^{\pi})^T W_{C} B_{\mathrm{cross}}^{\pi})$, and thus our goal becomes to show that $y^TKy \geq 0$ $\forall y \in \R_+^{n-1}$ (while allowing for sparsity in the selected constraints of course). 

\subsubsection{Understanding Crossing Indices Through Codes}

Now, we make a simple observation: a sufficient (but not necessary) condition for $K$ to satisfy $y^TKy \geq 0$ $\forall y \in \R_+^{n-1}$ is for \emph{every entry} in $K$ to be non-negative. Thus, it remains to understand exactly what the entries in $K$ look like: we start by looking at a simpler expression, namely $(B_{\mathrm{cross}}^{\pi})^T W_{C} B_{\mathrm{cross}}^{\pi}$. Indeed, in this matrix the $i,j$th entry can be written as
\[
	\left ( (B_{\mathrm{cross}}^{\pi})^T W_{C} B_{\mathrm{cross}}^{\pi} \right)_{i,j}= \sum_{c \in C} w_c \cdot \mathbf{1}[i \text{ crossing } c,\pi \wedge j \text{ crossing } c,\pi] = d_{C, \pi}(i,j),
	\]
	where we simply use $d_{C, \pi}(i,j)$ to denote the total weight of constraints that have both $i$ and $j$ as crossing indices under the permutation $\pi$.
	
Here comes the final, key technical lemma: we show that there is a matrix $G \in \F_2^{m \times n^2}$ such that \emph{for every} choice of $i, j$ and permutation $\pi$, there is a vector $z_{\pi, i, j} \in \F_2^{n^2}$ such that $(G z_{\pi, i, j})_{c}$ (i.e., the $c$th coordinate of the vector $(G z_{\pi, i, j})$) is exactly the indicator of whether or not $i, j$ are crossing indices for the constraint $c$ under permutation $\pi$. Thus, this value $d_{C, \pi}(i,j) = \left ( (B_{\mathrm{cross}}^{\pi})^T W_{C} B_{\mathrm{cross}}^{\pi} \right)_{i,j}$ can be written as the \emph{weighted hamming weight} of the vector $(G z_{\pi, i, j})$.

To construct this matrix $G$, we start with the generating matrix of our original XOR CSP $C$: this is the matrix $F \in \F_2^{m \times n}$, where for the constraint $c$ operating on $x_{u_1}, x_{u_2}, x_{u_3}, x_{u_4}$, there is a single row in the matrix $F$ corresponding to $c$, with $1$'s exactly in columns $u_1, u_2, u_3, u_4$. Now, the matrix $G$ is a type of \emph{tensor-product code} that is generated by $F \oplus F$. Explicitly, when we consider the linear space generated by $F$ (denoted $\mathrm{Im}(F)$), we want
\[
\mathrm{Im}(G) = \mathrm{Im}(F \oplus F) = \{ z_1 \cdot z_2: z_1, z_2 \in \mathrm{Im}(F) \},
\]
where $z_1 \cdot z_2$ refers to the \emph{entry-wise} multiplication of the two vectors. Note that $G$ will be expressible as a space of dimension $\leq O(n^2)$ as it essentially corresponds to a space of degree $2$ polynomials over $n$ variables (though see \cref{sec:generalProof} for a more thorough discussion). 

Now, recall that our goal is to show that for a fixed permutation $\pi$, as well as indices $i, j$, that we can find a vector in $\mathrm{Im}(G)$ whose $\ell$th coordinate is exactly the indicator of whether $i,j$ are crossing indices for the $\ell$th constraint under permutation $\pi$. For simplicity, let us suppose that the permutation $\pi$ is just the identity permutation; under this permutation, recall that an index $i$ is considered to be a crossing index for a constraint $c$ if and only if $i$ is in an active region of $c$. If we denote the variables in the constraint $c$ by $x_{u_1}, \dots x_{u_4}$, then $i$ is a crossing index under $\pi$ \emph{if and only if} $x_i$ is greater than or equal to an odd number of $x_{u_1}, x_{u_2}, x_{u_3}, x_{u_4}$: this is because the active regions are $[x_{u_1}, x_{u_2}]$ and $[x_{u_3}, x_{u_4}]$, and so $x_i$ must either be between $[x_{u_1}, x_{u_2}]$, or between $[x_{u_3}, x_{u_4}]$.

A priori, it may seem that this condition is completely arbitrary. However, it can be \emph{exactly} captured by our matrix $F$: indeed, let us consider the vector $e_{\leq i} \in \F_2^n$ which is $1$ in the first $i$ entries, and $0$ in the others. The $\ell$th constraint then performs an XOR on the variables $x_{u_1}, x_{u_2}, x_{u_3}, x_{u_4}$. Plugging this vector in, we obtain that
\[
(F e_{\leq i})_{\ell} = (e_{\leq i})_{u_1} \oplus (e_{\leq i})_{u_2} \oplus (e_{\leq i})_{u_3} \oplus (e_{\leq i})_{u_4},
\]
which is \emph{exactly} the indicator of whether $x_i$ is greater than or equal to an odd number of $x_{u_1}, x_{u_2}, x_{u_3}, x_{u_4}$ (since we are using the identity ordering of $x_1 \leq x_2 \leq \dots \leq x_n$).

Thus, the matrix $F$ alone captures when single indices are crossing a constraint. To generalize to \emph{pairs} of crossing indices, we then simply take the coordinate-wise product of whether indices $i,j$ are \emph{both} crossing a constraint $c$ under a permutation $\pi$. This is exactly what is captured by the code $\mathrm{Im}(G)$.  

\subsubsection{Creating Sparsifiers through Code Sparsification}

Because of this correspondence above, this means that $d_{C, \pi}(i,j)$ can be exactly written as the \emph{weighted hamming weight} of the \emph{codeword} $G z_{\pi, i, j}$, where $z_{\pi, i, j}$ is the particular vector we use to enforce 
\[
(G z_{\pi, i, j})_c = \mathbf{1}[i \text{ crossing } c,\pi \wedge j \text{ crossing } c,\pi] = d_{C, \pi}(i,j).
\]
To conclude then, we recall the work of Khanna, Putterman, and Sudan \cite{KPS24}, who showed that every code admits a sparsifier which preserves only a set of coordinates of size nearly-linear in the dimension of the code. In our case, the dimension is $O(n^2)$, so this means that there exists a re-weighted subset of coordinates $\hat{C}$, of size $\widetilde{O}(n^2 / \eps^2)$ which preserves the weight of every codeword to a $(1 \pm \eps)$ factor. In particular, this means for the CSP defined on the same re-weighted subset of constraints, it must be the case that 
\[
d_{\hat{C}, \pi}(i,j) \in (1 \pm \eps) d_{C, \pi}(i,j).
 \]
 
 By sparsifying this auxiliary code, we obtain a re-weighted subset of the constraints which satisfies that the parameter $d_{C, \pi}(i,j)$ is approximately preserved. When we revisit the matrices we created: $K = ( (B_{\mathrm{cross}}^{\pi})^T W_{\hat{C}} B_{\mathrm{cross}}^{\pi} - (1 - \eps)(B_{\mathrm{cross}}^{\pi})^T W_{C} B_{\mathrm{cross}}^{\pi})$, because the entries in $(B_{\mathrm{cross}}^{\pi})^T W_{\hat{C}} B_{\mathrm{cross}}^{\pi}$ are exactly $d_{\hat{C}, \pi}(i,j) $, and the entries in $(B_{\mathrm{cross}}^{\pi})^T W_{C} B_{\mathrm{cross}}^{\pi}$ are exactly $d_{C, \pi}(i,j)$, we see that the constraints $\hat{C}$ computed by the code sparsifier do indeed lead to a matrix with all non-negative entries as 
 \[
 d_{\hat{C}, \pi}(i,j) - (1 - \eps )d_{C, \pi}(i,j) \geq 0,
 \]
 and thus this matrix is non-negative on all non-negative vectors (and so too constitutes a spectral sparsifier by our previous discussions).
 
Likewise, observe that because the matrix $G$ we defined above is \emph{the same} for all permutations $\pi$ (we only need to change the \emph{vector} $e_{\leq i}$ above that we multiply the matrix by), this implies that the \emph{same} set of constraints will be a sparsifier across all choices of permutations, and thus we have a set of $\widetilde{O}(n^2 / \eps^2)$ constraints which is a spectral CSP sparsifier of our original instance. By generalizing this argument (and using the efficient constructions of code sparsifiers \cite{KPS24c}), we obtain \cref{thm:generalCSPintro}.

\subsubsection{Discussion}

We end the technical overview with some high-level remarks:

\begin{enumerate}
    \item Although we build on the framework of \cite{SY19}, the hypergraph sparsifiers in \cite{SY19} were of size $O(n^3 / \eps^2)$. Our refinement of their framework (specifically, the explicit connection with codes), allows for an improved sparsifier size of $\widetilde{O}(n^2 / \eps^2)$. 
    \item Likewise, the framework from \cite{SY19} is particular suited for sparsifying hypergraphs. Our framework reveals a much more general connection between spectral sparsification and sparsifying a type of \emph{tensor product} of the underlying object which holds \emph{across the entire CSP landscape}.
\item The method presented in this paper avoids any complex continuous machinery like chaining or matrix-chernoff, which have appeared in prior works on spectral sparsification \cite{KKTY21a, KKTY21b, JLLS23}. We view it is an interesting open question whether those techniques can be combined with ours to create nearly-linear size spectral CSP sparsifiers.
\end{enumerate}

\subsection{Organization}

In \cref{sec:prelim}, we re-introduce definitions from the introduction more formally, and include terminology that will be important in our proofs. In \cref{sec:XORproof}, we formally prove the sparsifiability of XOR constraints of arbitrary arity (i.e., a formal proof of the discussion held in the technical overview), and in \cref{sec:generalProof}, we prove the sparsifiability of all field-affine CSPs. Lastly, in \cref{sec:cheeger}, we prove a type of Cheeger inequality for our spectral notion of CSPs.

\section{Preliminaries}\label{sec:prelim}

In this work, we propose a notion of the energy of a CSP (analogous to the energy of a graph), and study the feasibility of constructing CSP sparsifiers which preserve this energy on all assignments.

\subsection{Energy Definitions}

To start, we recap the definition of a CSP:

\begin{definition}
    A CSP $C$ on $n$ variables and $m$ constraints is given by a predicate $P: \zo^r \rightarrow \zo$, along with $m$ ordered subsets of variables, denoted $S_1, \dots S_m$ where each $S_j \in [n]^r$. The $j$th constraint is then denoted by $P(x_{S_j})$, where the arguments in $P$ are understood to be $\{x_{\ell}: \ell \in S_j\}$. The CSP $C$ can also be accompanied by weights $w_c$ for each constraint $c \in C$. For a constraint $c \in C$, we also often use $c(x)$ to denote the value of the constraint when applied to the assignment of variables $x$ (even though $c$ may only be applied to a subset of $x$). 
\end{definition}

Next, we recall the notion of CSP sparsification. To do this, we first recall the definition of the value of a CSP:

\begin{definition}
    Let $C$ be a CSP on $n$ variables and $m$ constraints. For an assignment $x \in \zo^n$, the value of the assignment is
    \[
    |C(x)| = \sum_{j = 1}^m w_j \cdot P(x_{S_j}).
    \]
    In words, this is simply the weight of all satisfied constraints. 
\end{definition}

With this, we can introduce the notion of a CSP sparsifier:

\begin{definition}
    For a CSP $C$ on $n$ variables and $m$ constraints, a $(1 \pm \eps)$ sparsifier $\hat{C}$ of $C$ is a re-weighted CSP $\hat{C}$, with weights $\hat{w}_j: j \in [m]$ such that $\forall x \in \zo^n$:
    \[
    |\hat{C}(x)| \in (1 \pm \eps) \cdot |C(x)|.
    \]
    The sparsity of $\hat{C}$ is then given by $|\hat{w}|_0$.
\end{definition}

Observe that when the predicate $P$ is taken to be the 2-XOR function, the notion of CSP sparsification exactly captures that of cut-sparsification in graphs. However, in graphs there is also the more powerful notion of spectral sparsification:

\begin{definition}
    Let $G = (V, E)$ be a graph on $n$ vertices. For a vector $x \in [0,1]^V$, the energy of the graph is defined as
    \[
    Q_G(x) = \sum_{e = (u,v) \in E} (x_u - x_v)^2.
    \]
    A re-weighted sub-graph $\hat{G}$ is a $(1 \pm \eps)$ spectral sparsifier of $G$ if $\forall x \in [0,1]^V$:
    \[
    Q_{\hat{G}}(x) \in (1 \pm \eps)Q_{G}(x).
    \]
\end{definition}

With this, we are now ready to discuss a general notion of spectral sparsification for CSPs. To start, we will need the notion of the rounding of a vector $x \in [0,1]^n$:

\begin{definition}
    For a vector $x \in [0,1]^n$, and a value $\theta \in [0,1]$, the rounding of $x$ with respect to $\theta$ is the vector $x^{(\theta)}$ such that
    \[
    x^{(\theta)}_i = \mathbf{1}[x_i \geq \theta].
    \]
\end{definition}

This leads to a straightforward definition of the energy of a CSP:

\begin{definition}
    For a CSP $C$ on $n$ variables and $m$ constraints and a vector $x \in [0,1]^V$, the energy of $C$ is defined as 
    \[
    Q_C(x) = \sum_{c \in C} w_c \cdot \left(\Pr_{\theta \in [0,1]}[c(x^{(\theta)}) = 1] \right)^2.
    \]
\end{definition}

Our notion of spectral sparsification is then a natural continuation of this idea:

\begin{definition}
    For a CSP $C$ on $n$ variables and $m$ constraints, we say that a re-weighted sub-CSP $\hat{C}$ is a $(1 \pm \eps)$ spectral-sparsifier of $C$ if $\forall x \in [0,1]^V$
    \[
    Q_{\hat{C}}(x) \in (1 \pm \eps)Q_{C}(x).
    \]
\end{definition}

A few remarks are in order: first, when a vector $x \in \zo^n$, the energy $Q_C(x)$ is exactly the same as the corresponding value of the CSP on $x$. This is because over a random choice of $\theta$, the rounding of the vector is exactly the same as the vector itself. Thus, $\Pr_{\theta \in [0,1]}[c(x^{(\theta)}) = 1]$ is either $0$ or $1$, and taking the square does not alter the value. Second, one can check that when we use a $2$-XOR constraint on variables $x_u, x_v$, the probability over a random $\theta$ that the constraint is satisfied is exactly $|x_u - x_v|$ (the probability that $\theta$ lands in the interval $(x_u, x_v)$). Squaring this quantity then yields exactly the notion of energy in a graph. 

\subsection{Copositivity and Permutations}

In this section, we will now recap some primitives regarding copositivity and permutations that will be useful in our proofs of spectral CSP sparsification. 
\begin{definition}
    A symmetric matrix $A \in \R^{n \times n}$ is said to be \emph{copositive} if $x^T A x \geq 0$, for every $x \in \R_+^{n}$.
\end{definition}

Copositivity will arise when we study orderings associated with a vector $x \in \R^n$:

\begin{definition}
    Let $\pi$ be a permutation on $[n]$. Then, the set $[0,1]^{\pi} \subseteq [0,1]^n$ is the set of vectors such that $x(\pi_1) \geq x(\pi_2) \geq \dots \geq x(\pi_n)$.
\end{definition}

Using this, we can define a restricted notion of copositivity:

\begin{definition}
    For a symmetric matrix $A \in \R^{n \times n}$, and a permutation $\pi$ on $[n]$, we say that $A$ is $[0,1]^{\pi}$-copositive if $x^T A x \geq 0$ for all $x \in [0,1]^{\pi}$.
\end{definition}

We will also use a few special matrices: 

\begin{definition}\label{def:specialMatrices}
\begin{enumerate}
    \item For permutation $\pi$, we define the permutation matrix $P_{\pi} \in \R^{n \times n}$, where $P_{\pi}(i,j) = 1$ if $j = \pi_i$, and otherwise is $0$.
    \item For an integer $n$, we also define the matrix $J \in \R^{(n-1) \times n}$, where $J(i,i) = 1$, $J(i, i+1) = -1$, and otherwise is $0$.
\end{enumerate}
\end{definition}

Finally, with these definitions, \cite{SY19} created the following characterization of copositivity:

\begin{lemma}[Lemma 3.3 in \cite{SY19}]\label{lem:PSDtoCopos}
    Let $A \in \R^{n \times n}$ be a symmetric matrix, such that $A \cdot \mathbf{1} = 0$, and let $\pi$ be a permutation on $[n]$. Then, $A$ is $[0,1]^{\pi}$-copositive if and only if $A$ is of the form $P_{\pi}^T J^T K J P_{\pi}$ for some copositive matrix $K \in \R^{(n-1) \times (n-1)}$.
\end{lemma}

\section{Spectral Sparsification of XOR CSPs}\label{sec:XORproof}

In this section, we now present a proof of spectral sparsifier for CSPs on XOR constraints. Specifically, we will show the following theorem:

\begin{theorem}\label{thm:XORmain}
    Let $C$ be an XOR CSP on $n$ variables. There is a (probabilistic) polynomial time algorithm which produces (with high probability) a $(1 \pm \eps)$ spectral sparsifier $\hat{C}$ of $C$ such that $\hat{C}$ retains only $\widetilde{O}(n^2 / \eps^2)$ re-weighted constraints.
\end{theorem}

\subsection{Related Matrices}

To prove this theorem, we will introduce some notation and terminology specific to XOR CSPs. To start, recall that the energy of the CSP $C$ is given by 

\[
    Q_C(x) = \sum_{c \in C} w_c \cdot \left(\Pr_{\theta \in [0,1]}[c(x^{(\theta)}) = 1] \right)^2.
\]

When each constraint $c$ is an XOR constraint applied to a subset of the variables, we can give an exact expression for what $\left(\Pr_{\theta \in [0,1]}[c(x^{(\theta)}) = 1] \right)^2$ evaluates to. Indeed, let $S \subseteq [n]$ denote the set of indices on which the XOR constraint is applied, and let $x_{o_{i}}$ denote the $i$th smallest element in $\{x_{\ell}: \ell \in S\}$. Then,
\[
\left(\Pr_{\theta \in [0,1]}[c(x^{(\theta)}) = 1] \right)^2 = (x_{o_{|S|}} - x_{o_{|S|-1}} + x_{o_{|S|-2}} - x_{o_{|S|-3}} + \dots x_{o_2} - x_{o_1})^2.
\]

In particular, observe that once an ordering $\pi$ on $[n]$ is fixed, the above expression is simply a quadratic form. I.e., let us suppose that the vector $x$ is such that $x_1 \leq x_2 \leq x_3 \leq \dots \leq x_n$, and consider an XOR on the variables $x_1, x_2, x_3, x_4$. Then, the above expression for energy evaluates exactly to $(x_4 - x_3 + x_2 - x_1)^2$.

Going forward, let us consider a fixed permutation $\pi$ (i.e., an ordering of the $x_i$'s). Let us denote $u_1, u_2, \dots u_{|S|}$ to be the resulting ordering of the elements of $S$. I.e., $x_{u_1} \leq x_{u_2} \leq x_{u_3} \leq \dots \leq x_{u_{|S|}}$.

To summarize, we have the following claim:

\begin{claim}\label{clm:fixedOrderingExp}
    Let $C$ be an XOR CSP on $n$ variables, where each constraint acts on a subset of these variables of even size. Let $\pi$ be a fixed ordering of $[n]$, and let $x$ be any vector in $[0,1]^{\pi}$. For a constraint $c$, let $S_c \subseteq [n]$ denote the indices of the variables that $c$ operates on, and let 
    $x_{u_{c, 1}} \leq x_{u_{c, 2}} \leq x_{u_{c, 3}} \leq \dots \leq x_{u_{c, |S_c|}}$ be the resulting ordering of the elements of $S_c$ under $\pi$. Then,
    \[
    \left(\Pr_{\theta \in [0,1]}[c(x^{(\theta)}) = 1] \right)^2 = (x_{u_{c, |S|}} - x_{u_{c, |S|-1}} + \dots + x_{u_{c, 2}} - x_{u_{c, 1}})^2.
    \]
\end{claim}
 
 Now, observe that we can write the above expression as the sum of differences between consecutive pairs. We denote the values between these consecutive elements as the \emph{active regions}. More explicitly:

\begin{definition}
    For a fixed ordering $\pi$ on $[n]$, and an XOR constraint $c$ operating on an even number of ordered elements $x_{u_1} \leq x_{u_2} \leq x_{u_3} \leq \dots \leq x_{u_{|S|}}$, we say the active regions are 
    \[
    \bigcup_{j \in \N} [x_{u_{2 \cdot j-1}}, x_{u_{2 \cdot j}}].
    \]
\end{definition}

For an index $i \in [n-1]$, let us consider the bisector between $x_{\pi_i}$ and $x_{\pi_{i+1}}$ (i.e., between the $i$th and $i+1$st smallest elements in the ordering $\pi$). This bisector takes on value $(x_{\pi_i} + x_{\pi_{i+1}}) / 2$. 

\begin{definition}
For $i \in [n-1]$, we say that $i$ is a crossing index with respect to the ordering $\pi$ and constraint $c$ if the bisector $(x_{\pi_i} + x_{\pi_{i+1}}) / 2$ is in an active region of the constraint $c$ with ordering $\pi$.
\end{definition}

We also generalize this definition to a pair of indices $i,j \in [n-1]$:

\begin{definition}
    We say that pairs of indices $i,j \in \binom{[n-1]}{2}$ are crossing indices with respect to the constraint $c$, ordering $\pi$ if both $i, j$ individually are crossing indices with $c, \pi$.
\end{definition}

\begin{definition}\label{def:NumberCrossing}
	For an XOR CSP $C$, a permutation $\pi$, and an index $i$, we let
    \[
	 d_{C, \pi}(i) = \sum_{c \in C} w_c \cdot \mathbf{1}[i \text{ crossing } c,\pi].
	\]
    For a pair of
    indices $i, j$, we let $d_{C, \pi}(i,j)$ be the total weight of constraints for which $i,j$ are both crossing under $\pi$. I.e.,
	\[
	 d_{C, \pi}(i,j) = \sum_{c \in C} w_c \cdot \mathbf{1}[i \text{ crossing } c,\pi \wedge j \text{ crossing } c,\pi].
	\]
\end{definition}

With this, we now introduce some auxiliary matrices:

\begin{definition}\label{def:incMatrixXOR}
	Let $C$ be an XOR CSP on $n$ variables, and $m$ constraints, where each XOR constraint operates on an even number of variables, and let $\pi$ be an ordering of the variables $x_1, \dots x_n$. Let $c$ be a single constraint, and let $x_{u_1} \leq x_{u_2} \leq \dots \leq x_{u_{|S|}}$ be the ordered variables on which $c$ operates. We say that $B^{\text{inc}}_{C, \pi} \in \R^{m \times n}$ is the \emph{incidence matrix} for $C$, where for $c \in C$, $i \in [n]$, $B^{\text{inc}}_{C, \pi}(c, i) = 0$ if $x_i$ is not present in the constraint $c$, $B^{\text{inc}}_{C, \pi}(c, i) = -1$ if $x_i = x_{u_j}$ and $j = 0 \mod 2$ (i.e., an upper border of part of the active region), and otherwise $B^{\text{inc}}_{C, \pi}(c, i) = 1$ (when $x_i$ is a lower border of part of the active region). 
\end{definition}

\begin{remark}\label{rmk:incidence}
	The incidence matrix $B^{\text{inc}}_{C, \pi}$ is constructed such that for $x \in [0,1]^{\pi}$, $B^{\text{inc}}_{C, \pi}(x)_c = -(x_{u_{|S|}} - x_{u_{|S|-1}} + \dots + x_{u_2} - x_{u_1})$.
\end{remark}

\begin{claim}\label{clm:quadraticEquivalence}
	Let $C$ be an XOR CSP as in \cref{def:incMatrixXOR}, and let $W_C$ be the diagonal matrix in $\R^{m \times m}$ such that $W_C(c, c) = w_c$ (i.e., the diagonal matrix of weights of constraints). Then for $x \in [0,1]^{\pi}$
	\[
	x^T (B^{\text{inc}}_{C, \pi})^T W_C B^{\text{inc}}_{C, \pi} x = Q_C(x).
	\]
\end{claim}

\begin{proof}
	We have that 
	\[
	x^T (B^{\text{inc}}_{C, \pi})^T W_C B^{\text{inc}}_{C, \pi} x = \sum_{c \in C} w_c \cdot (B^{\text{inc}}_{C, \pi}(x)_c)^2 = \sum_{c \in C} w_c \cdot (x_{u_{c, |S|}} - x_{u_{c, |S|-1}} + \dots + x_{u_{c, 2}} - x_{u_{c, 1}})^2 = Q_C(x),
	\]
	where the second to last equality holds because $B^{\text{inc}}_{C, \pi}(x)_c$ is exactly the expression $-(x_{u_{c, |S|}} - x_{u_{c, |S|-1}} + \dots + x_{u_{c, 2}} - x_{u_{c, 1}})$ via \cref{rmk:incidence}, and the final equality holds because of \cref{clm:fixedOrderingExp}. Because we square the expression, the negative sign is inconsequential.
\end{proof}

Now, we introduce another auxiliary matrix:

\begin{definition}\label{def:crossMatrixXOR}
    Let $C$ be an XOR CSP as in \cref{def:incMatrixXOR}. We define the matrix $B^{\text{cross}}_{C, \pi} \in \R^{m \times n-1}$ such that $B^{\text{cross}}_{C, \pi}(c, i) = 1$ if $i$ is a crossing index for $c$ under $\pi$, and otherwise $B^{\text{cross}}_{C, \pi}(c, i) = 0$.
\end{definition}

Next, we will re-write $B^{\text{inc}}_{C, \pi}$ in a more convenient way:

\begin{claim}\label{clm:convenientRep}
	$B^{\text{inc}}_{C, \pi} = B^{\text{cross}}_{C, \pi} J P_{\pi}$, for $B^{\text{inc}}_{C, \pi}$ as defined in \cref{def:incMatrixXOR}, $B^{\text{cross}}_{C, \pi}$ as defined in \cref{def:crossMatrixXOR}, and $J, P_{\pi}$ as defined in  \cref{def:specialMatrices}.
\end{claim}

\begin{proof}
	Let us first prove this when $\pi = \mathrm{id}_n$. Let $c$ be a given XOR constraint, operating on elements $\{u_1, \dots u_r \} \subseteq [n]$. Observe that 
	\[
	(B^{\text{cross}}_{C, \mathrm{id}_n} J)_{c, i} = \sum_{\ell=  1}^{n-1} (B^{\text{cross}}_{C, \mathrm{id}_n})_{c, \ell} \cdot J_{\ell, i} = -(B^{\text{cross}}_{C, \mathrm{id}_n})_{c, i-1} + (B^{\text{cross}}_{C, \mathrm{id}_n})_{c, i}.
	\]
	
	So, all that remains is to understand this quantity in different settings:
	\begin{enumerate}
		\item When $i-1, i$ are both crossing elements for $c$ under $\mathrm{id}_n$. Then, $(B^{\text{cross}}_{C, \mathrm{id}_n} J)_{c, i} = 0$. Likewise, if $i-1$, $i$ are both crossing elements, then the active region completely contains $[x_{i-1}, x_i], [x_i, x_{i+1}]$. Hence, the incidence matrix would also place a coefficient of $0$ on $x_i$.
		\item When neither of $i-1, i$ are crossing elements for $c$ under $\mathrm{id}_n$. Then, $(B^{\text{cross}}_{C, \mathrm{id}_n} J)_{c, i} = 0$. Likewise, if $i-1$, $i$ are both crossing elements, then the active region does not contain $[x_{i-1}, x_i]$ or $[x_i, x_{i+1}]$. Hence, the incidence matrix should place a coefficient of $0$ on $x_i$.
		\item Now, let us suppose that only $i-1$ is a crossing element. Then, the above expression yields $(B^{\text{cross}}_{C, \mathrm{id}_n} J)_{c, i} = -1$. However, observe that if only $i-1$ is a crossing element, this means that $[x_{i-1}, x_i]$ is in the active region, while $[x_i, x_{i+1}]$ is not in the active region, and thus $x_i$ is the upper border of part of the active region, and the incidence matrix would put a $-1$ corresponding to $x_i$.
		\item Finally, let us suppose that only $i$ is a crossing element. Then, the above expression yields $(B^{\text{cross}}_{C, \mathrm{id}_n} J)_{c, i} = 1$. However, observe that if only $i$ is a crossing element, this means that $[x_{i-1}, x_i]$ is not in the active region, while $[x_i, x_{i+1}]$ is in the active region, and thus $x_i$ is the lower border of part of the active region. So, the incidence matrix would place a $1$ in this coordinate.
	\end{enumerate}
	
	Now, for general permutations, $P_{\pi}$ will simply compensate for the fact that $i$ is a crossing index only in the permuted order corresponding to $\pi$, and thus the above correspondence is unchanged. So, $B^{\text{inc}}_{C, \pi} = B^{\text{cross}}_{C, \pi} J P_{\pi}$.
\end{proof}

Finally, we have the following simple observation about $B^{\text{cross}}_{C, \pi}$:

\begin{claim}\label{clm:crossingNumberCrossingMatrix}
	Let $C$ be an XOR CSP as in \cref{def:incMatrixXOR}. Then, for $i, j \in [n-1]$ and a permutation $\pi$, we have that $((B^{\text{cross}}_{C, \pi})^T W_C B^{\text{cross}}_{C, \pi})_{i,j} = d_{C, \pi}(i,j)$.
\end{claim}

\begin{proof}
	Observe that 
	\[
	((B^{\text{cross}}_{C, \pi})^T W_C B^{\text{cross}}_{C, \pi})_{i,j} = \sum_{c \in C} w_c \cdot \mathbf{1}[i \text{ crossing } c,\pi \wedge j \text{ crossing } c,\pi] = d_{C, \pi}(i,j),
	\]
    as per \cref{def:NumberCrossing}.
\end{proof}

With these notions established, we are now ready to study the sparsification of these XOR CSPs.

\subsection{Reducing to Code Sparsification}

In this section, we show how to spectrally sparsify CSPs by using the framework of code sparsification \cite{KPS24, KPS24c}. To start, we establish the following sufficient condition for spectral sparsification:

\begin{claim}\label{clm:copositiveMatrix}
	Let $C$ be an XOR CSP as in \cref{def:incMatrixXOR}. Then $\hat{C}$ (with corresponding weight matrix $W_{\hat{C}}$) is a $(1 \pm \eps)$ spectral sparsifier of $C$ if $\forall \pi$, both
	\[
	(B^{\text{cross}}_{C, \pi})^T (W_{\hat{C}} - (1 - \eps)W_C) B^{\text{cross}}_{C, \pi},
	\]
	and 
	\[
	(B^{\text{cross}}_{C, \pi})^T ((1 + \eps)W_C - W_{\hat{C}} ) B^{\text{cross}}_{C, \pi} 
	\]
	are copositive for vectors in $\R^{n-1}_+$.
\end{claim}

\begin{proof}
Indeed, let $\pi$ be a fixed permutation, and let $\hat{C}$ be created by sampling constraint in accordance with the aforementioned weighting scheme. Our goal is to show that for $x \in [0,1]^{\pi}$
\[
(1 - \eps)Q_C(x) \leq Q_{\hat{C}}(x) \leq (1 + \eps)Q_C(x).
\]
By \cref{clm:quadraticEquivalence}, this is equivalent to showing that for $x \in [0,1]^{\pi}$
\[
(1 - \eps) x^T (B^{\text{inc}}_{C, \pi})^T W_C B^{\text{inc}}_{C, \pi} x  \leq x^T (B^{\text{inc}}_{C, \pi})^T W_{\hat{C}} B^{\text{inc}}_{C, \pi} x \leq (1 + \eps) x^T (B^{\text{inc}}_{C, \pi})^T W_C B^{\text{inc}}_{C, \pi} x,
\]
where $W_{\hat{C}} $ is now the diagonal matrix of weights corresponding to the sparsified CSP $\hat{C}$.

Now, we also invoke \cref{clm:convenientRep}. This implies the above copositivity is in fact equivalent to showing that 
\[
(1 - \eps) x^T P_{\pi}^{T} J^T (B^{\text{cross}}_{C, \pi})^T W_C B^{\text{cross}}_{C, \pi} J P_{\pi} x  \leq x^T P_{\pi}^{T} J^T (B^{\text{cross}}_{C, \pi})^T W_{\hat{C}} B^{\text{cross}}_{C, \pi} J P_{\pi} x 
\]
\[
\leq (1 + \eps) x^T P_{\pi}^{T} J^T (B^{\text{cross}}_{C, \pi})^T W_C B^{\text{cross}}_{C, \pi} J P_{\pi} x,
\]
for $x \in [0,1]^{\pi}$.

Now, let us focus only on one side of the above expression: namely to show that \[
(1 - \eps) x^T P_{\pi}^{T} J^T (B^{\text{cross}}_{C, \pi})^T W_C B^{\text{cross}}_{C, \pi} J P_{\pi} x  \leq x^T P_{\pi}^{T} J^T (B^{\text{cross}}_{C, \pi})^T W_{\hat{C}} B^{\text{cross}}_{C, \pi} J P_{\pi} x
\]for $x \in [0,1]^{\pi}$. This is equivalent to showing 
\[
x^T P_{\pi}^{T} J^T (B^{\text{cross}}_{C, \pi})^T (W_{\hat{C}} - (1 - \eps)W_C) B^{\text{cross}}_{C, \pi} J P_{\pi} x \geq 0.
\]

Because this is already in the form of \cref{lem:PSDtoCopos}, in order to prove the above is copositive, we must only show that 
\[
(B^{\text{cross}}_{C, \pi})^T (W_{\hat{C}} - (1 - \eps)W_C) B^{\text{cross}}_{C, \pi}
\]
is copositive for $x \in \R^{n-1}_+$, as we desire. Analogous reasoning shows the claim for the $(1 + \eps)$ side of the expression, yielding our claim. 
\end{proof}

Now, we recall the notion of a generating matrix:

\begin{definition}\label{def:genMatrixXOR}
    Let $C$ be an XOR CSP on $n$ variables with $m$ constraints. Let $S_c \subseteq [n]$ denote the set of variables on which constraint $c$ is operating. The corresponding generating matrix $G \in \F_2^{m \times n}$ is defined such that $G_{c, i} = \mathbf{1}[i \in S_c]$.
\end{definition}

We also use the following vector:

\begin{definition}
	Let $\pi$ be an ordering on $[n]$, and let $i \in [n]$ be an index. We say that $z_{\pi, > i}$ is the vector in $\zo^n$ such that $z_{\ell} = 1$ if and only if $\pi_{\ell} < \pi_i$.
\end{definition}

\begin{claim}\label{clm:crossingEquivalence}
	Let $C$ be an XOR CSP on $n$ variables, $m$ constraints, and let $G$ be its canonical generating matrix. Then, for a fixed permutation $\pi$ on $n$, we have that 
	\[
	d_{C, \pi}(i) = \wt(G z_{\pi, > i}).
	\]
\end{claim}

\begin{proof}
	Let us consider an XOR constraint $c \in C$ acting on a subset $S$ of indices. Recall that for an ordering $\pi$, the index $i$ is \say{crossing} the constraint $c$ if the sector $(x_{\pi_i}, x_{\pi_{i+1}})$ lies in one of the active regions of $c$. Importantly however, recall that the active regions of $c$ are alternating. That is to say, the region between the largest element in $x_{\ell}: \ell \in S_c$ and the second largest is active, while the region between the second and third largest is not active. Again though, between the third largest and fourth largest is active, while between the fourth and fifth is not. This means that the sector $(x_{\pi_i}, x_{\pi_{i+1}})$ is in an active region \emph{if and only if} there are an odd number of $x_{\ell}: \ell \in S$ for which $x_{\ell} > x_{\pi_i}$, and thus $i$ is a crossing index for $c, \pi$ if and only if this happens.
	
	Now, for the constraint $c$ let $S_c \subseteq [n]$ denote the variables on which $c$ operates. Observe that $(G z_{\pi, > i})_c$ is exactly an XOR of all $(z_{\pi, > i})_{\ell}: \ell \in S_c$. But, an entry $(z_{\pi, > i})_{\ell}$ is $1$ if and only if $\pi_{\ell} < \pi_i$, correspondingly if $x_{\pi_{\ell}} > x_{\pi_i}$, and therefore is larger than $x_{\pi_i}$ in the ordering. Thus, this XOR evaluates to $1$ iff there are an odd number of $x_{\ell}: \ell \in S_c$ for which $x_{\ell} > x_{\pi_i}$.
	
	By summing across all constraints $c \in C$ (and weighting appropriately) we get the desired claim. 
\end{proof}

As an immediate corollary, we also establish the following for pairs $i,j$:

\begin{claim}\label{clm:crossingProduct}
	Let $c$ be an XOR constraint on a subset $S_c \subseteq [n]$ of the variables, let $\pi$ be a permutation on $n$, let $i \neq j \in [n-1]$ be two indices, and let $G_c$ denote the row of the canonical generating matrix $G$ corresponding to constraint $c$. Then, 
	\[
	\mathbf{1}[i,j \text{ crossing } c, \pi] = (G z_{\pi, > i})_c \cdot (G z_{\pi, > j})_c.
	\]
\end{claim}

\begin{proof}
	\cref{clm:crossingEquivalence} establishes that $i$ is crossing $c$ with $\pi$ if and only if $G z_{\pi, > i} = 1$. Thus, $i, j$ are \emph{both} crossing $c, \pi$ if and only if the \emph{product} $G z_{\pi, > i} \cdot G z_{\pi, > j} = 1$.
\end{proof}

So, we now define a new coding theoretic object to capture the above notion. Specifically, observe that we are taking the AND of two XORs. Over $\F_2$, the XORs are simply linear functions, while the AND is taking the product of these linear functions. For a constraint operating on a fixed set $S$, this function becomes 
\[
(\oplus_{i \in S} x_i) \cdot (\oplus_{i \in S} y_i),
\]
where the $x$ would correspond to $z_{\pi, > i}$, and $y$ would correspond to $z_{\pi, > j}$. Thus, this is essentially a degree $2$ polynomial over $2n$ variables (though there is some relationship between $x, y$). Finally, we take advantage of the fact that we can replace each variable $x_a \cdot y_{b}$ with an \emph{auxiliary} variable, which we denote by $q_{ab}$. Once we perform this simplification, we effectively have a \emph{linear} constraint over $\F_2$, which is operating on a universe of $n^2$ variables:

\begin{definition}\label{def:liftedGenerating}
For a constraint $c$, we let $\Phi(c)$ denote this lifted version of the constraint to the universe of $n^2$ variables. We let $\Phi(C)$ denote the result of lifting the entire CSP, and we let $\Phi(G) \in \F_2^{m \times n^2}$ denote the result of lifting the canonical generating matrix. 
\end{definition}

Next, recall that by \cref{clm:crossingProduct}, for a given constraint $c$, we have that $i,j$ are crossing $c$ for permutation $\pi$ if and only if 
\[
c(z_{\pi, > i}) \cdot c(z_{\pi, > i}) = 1.
\]

This yields the following claim:

\begin{claim}\label{clm:preserveAndWeights}
	Let $C$ be an XOR CSP. Then $\hat{C}$ is a $(1 \pm \eps)$ spectral sparsifier of $C$ if $\forall \pi$, and $\forall i, j \in [n-1]$:
	\[
	\wt(Gz_{\pi, > i} \cdot Gz_{\pi, > i}) \in (1 \pm \eps) \wt(\hat{G}z_{\pi, > i} \cdot \hat{G}z_{\pi, > i}),
	\]
	where $G$ is the canonical generating matrix of $C$, and $\hat{G}$ is the canonical generating matrix of $\hat{C}$.
\end{claim}

\begin{proof}
	Recall that by \cref{clm:crossingProduct}, for a permutation $\pi$, and indices $i, j$, we have that 
	\[
	\wt(Gz_{\pi, > i} \cdot Gz_{\pi, > i}) = \sum_{c \in C} w_c \cdot 
	(G z_{\pi, > i})_c \cdot (G z_{\pi, > j})_c = \sum_{c \in C} w_c \cdot \mathbf{1}[i,j \text{ crossing } c, \pi] 
	\]
	\[
	= d_{C, \pi}(i,j). 
	\]
	
	Likewise, when we use the canonical generating matrix for $\hat{C}$, we get that 
	\[
	\wt(\hat{G}z_{\pi, > i} \cdot \hat{G}z_{\pi, > i}) = d_{\hat{C}, \pi}(i,j).
	\]
	
	Finally, recall that by \cref{clm:crossingNumberCrossingMatrix}, for a fixed permutation $\pi$, we have that 
	\[
	(B^{\text{cross}}_{C, \pi})^T W_C B^{\text{cross}}_{C, \pi} = d_{C, \pi}(i,j),
	\]
	and 
	\[
	(B^{\text{cross}}_{C, \pi})^T W_{\hat{C}} B^{\text{cross}}_{C, \pi} = d_{\hat{C}, \pi}(i,j).
	\]
	
	Thus, if $\forall \pi$, and $\forall i, j \in [n-1]$
	\[
	\wt(Gz_{\pi, > i} \cdot Gz_{\pi, > i}) \in (1 \pm \eps) \wt(\hat{G}z_{\pi, > i} \cdot \hat{G}z_{\pi, > i}),
	\]
	this means that all entries in 
	\[(B^{\text{cross}}_{C, \pi})^T (W_{\hat{C}} - (1 - \eps)W_C) (B^{\text{cross}}_{C, \pi})^T\]
	and 
	\[
	(B^{\text{cross}}_{C, \pi})^T ((1 + \eps)W_C - W_{\hat{C}} ) B^{\text{cross}}_{C, \pi} 
	\]
	are positive, and thus the matrices are copositive for vectors in $\R^{n-1}_+$. By \cref{clm:copositiveMatrix}, this then means that $\hat{C}$ is a $(1 \pm \eps)$ spectral sparsifier of $C$.
\end{proof}

Finally, this means it suffices to find a set of weights $w_{\hat{C}}$ such that for $\forall \pi$ and $\forall i, j \in [n-1]$:
\[
\wt(Gz_{\pi, > i} \cdot Gz_{\pi, > i}) \in (1 \pm \eps) \wt(\hat{G}z_{\pi, > i} \cdot \hat{G}z_{\pi, > i}).
\]
To do this, we will once again use the lifted version of the generating matrices, and perform \say{code sparsification} on these matrices. We recall the definition of a code sparsifier from \cite{KPS24}:

\begin{definition}
	Let $G \in \F_2^{m \times n}$ be a generating matrix. We say that $\hat{G}$ is a $(1 \pm \eps)$-code sparsifier for $G$ if the rows of $\hat{G}$ are a re-weighted subset of the rows of $G$, and $\forall x \in \F_2^n$:
	\[
	\wt(\hat{G}x) \in (1 \pm \eps) \wt(Gx).
	\]
	Here, we use $\wt$ to mean the \emph{weighted hamming weight}, in the sense that certain coordinates may be weighted more in their respective code. 
\end{definition}

The work of \cite{KPS24} showed the following:

\begin{theorem}\cite{KPS24}\label{thm:codeSparsification}
	For any generating matrix $G \in \F_2^{m \times n}$, there is a $(1 \pm \eps)$ code-sparsifier $\hat{G}$ for $G$, such that $\hat{G} \in \F_2^{m' \times n}$, where $m' = \widetilde{O}(n / \eps^2)$.
\end{theorem}

Further, \cite{KPS24c} showed that such a code sparsifier can be computed in \emph{polynomial time} in $m, n, \eps$ (with high probability). We are now ready to provide the main proof of this section:

\begin{proof}[Proof of \cref{thm:XORmain}]
Given the XOR CSP $C$, on $n$ variables, we first add a dummy variable $x_{n+1}$. If any constraint $c$ operates on an odd number of variables, we augment the constraint $c$ to also have an XOR with $x_{n+1}$. In this way, every constraint WLOG has even arity. Next, we construct the corresponding generating matrix $G$, as well as the \emph{lifted} generating matrix $\Phi(G)$ in accordance with \cref{def:liftedGenerating}. 
	
	Next, we invoke \cref{thm:codeSparsification} to create a $(1 \pm \eps)$ code-sparsifier of $\Phi(G)$, which we denote by $\widehat{\Phi(G)}$. Note that because the dimension of $\Phi(G)$ is $n^2$, the number of rows (coordinates) preserved in $\widehat{\Phi(G)}$ is $\widetilde{O}(n^2 / \eps^2)$.
	
	Further, observe that every row of $\widehat{\Phi(G)}$ actually corresponds to a row of $\Phi(G)$. In particular, $\widehat{\Phi(G)} = \Phi(\widehat{G})$, where $\hat{G}$ is simply the generating matrix where each coordinate $j \in [m]$ of $G$ has the same weight as the coordinate in $\widehat{\Phi(G)}$, but where we keep row $G_j$ instead of $\Phi(G)_j$. Note that this follows because the lifting operation is applied individually to each row of $G$, and thus a collection of lifted rows is equivalent to lifting the collection of (unlifted) rows. Going forward, we will denote the CSP associated with $\hat{G}$ by $\hat{C}$.
	
	Next, for any $\pi$, and any $i, j \in [n-1]$, consider the vector $x_{\pi, i, j} \in \zo^{n^2}$ where 
     $x_{\pi, i, j} = z_{\pi, > i} \otimes z_{\pi, > j}$. Observe that for any constraint $c$, (and the lifted version $\Phi(c) \in \F_2^{n^2}$), we have
	\[
	\Phi(c) x_{\pi, i, j} = (G z_{\pi, > i})_c \cdot  (G z_{\pi, > j})_c = \mathbf{1}[i,j \text{ crossing } c, \pi].
	\]
	Thus, for this vector $x_{\pi, i, j}$, we have that 
	\[
	\wt(\Phi(G) x_{\pi, i, j}) = \sum_{c \in C} w_c  \mathbf{1}[i,j \text{ crossing } c, \pi] =\wt(Gz_{\pi, > i} \cdot Gz_{\pi, > i}).
	\]
	and likewise
	\[
	\wt(\Phi(\widehat{G}) x_{\pi, i, j}) = \wt(\hat{G}z_{\pi, > i} \cdot \hat{G}z_{\pi, > i}),
	\]
	
	Because $\Phi(\hat{G})$ is a $(1 \pm \eps)$ code-sparsifier of $\Phi(\hat{G})$, and the message $x_{\pi, i, j}$ is fixed, this must mean that 
	\[
	\wt(\hat{G}z_{\pi, > i} \cdot \hat{G}z_{\pi, > i}) \in (1 \pm \eps) \wt(Gz_{\pi, > i} \cdot Gz_{\pi, > i}).
	\]
	
	Thus, we have constructed a CSP $\hat{C}$ such that $\forall \pi$, $\forall i, j \in [n-1]$, 
	\[
	\wt(Gz_{\pi, > i} \cdot Gz_{\pi, > i}) \in (1 \pm \eps) \wt(\hat{G}z_{\pi, > i} \cdot \hat{G}z_{\pi, > i}),
	\]
	and we conclude the spectral sparsification via \cref{clm:preserveAndWeights}.

    Note that even though we augmented the CSP $C$ with the extra variable $x_{n+1}$ to ensure the even arity, the sparsifier we obtain is still a sparsifier for the original CSP. In particular, we can set $x_{n+1} = 0$, and then the energy of the augmented CSP is exactly the same as the un-augmented CSP. 
	
	The running time / probability bounds follow by using tools for efficient code sparsification from \cite{KPS24c}. 
\end{proof}

The notation and techniques in this section were simplified for the exposition of spectrally sparsifying XOR CSPs. In the following section, we show a more general approach for sparsifying any CSP which can be written as a linear equation over a field. 

\section{Spectral Sparsification of Field-Affine CSPs}\label{sec:generalProof}

To start, we will generalize some of the preliminaries from the previous section to arbitrary CSP predicates. Then, we will show how a framework parallel to the one above can be used to show the following theorem:

\begin{theorem}\label{thm:generalCSP}
    Let $C$ be any CSP on $n$ variables and $m$ constraints using a predicate $P: \zo^r \rightarrow \zo$ such that $P(y) = 1 \iff \mathbf{1}[a_i y_i \neq b \mod p]$, for some prime $p$. Then, there is a randomized, polynomial time algorithm which with high probability computes a $(1 \pm \eps)$ spectral-sparsifier of $C$ that retains only $\widetilde{O}(n^2 \log^2(p) / \eps^2)$ re-weighted constraints.
\end{theorem}

\subsection{General Spectral CSP Notions}

First, we formalize the definition of a field-affine CSP:

\begin{definition}\label{def:fieldaffine}
    We say that a CSP $C$ on $n$ variables and $m$ constraints using a predicate $P: \zo^r \rightarrow \zo$ is field-affine if there exists a prime $p$ and coefficients $a_i, b$ such that $P(y) = 1 \iff \mathbf{1}[a_i y_i \neq b \mod p]$.
\end{definition}

\begin{remark}\label{rmk:zeroPredicate}
Note that we will be assuming that our predicates $P$ also satisfy that $P(1^r) = P(0^r) = 0$. This assumption is actually \emph{without} loss of generality. Indeed, because $P(x) = \mathbf{1}[\sum_i a_i x_i \neq b \mod p]$, we can re-write this using auxiliary random variables $x_{r+1}, x_{r+2}$ to make a predicate $P'(x, x_{r+1}, x_{r+2}) = \mathbf{1}[\sum_i a_i x_i - b \cdot x_{r+1} + x_{r+2} \cdot (\sum_i a_i - b) \neq 0 \mod p]$. When $x_{r+1} = 1, x_{r+2} = 0$, then 
\[
P'(x, 1, 0) = \mathbf{1}[\sum_i a_i x_i \neq b \mod p],
\]
and likewise
\[
P'(1^r, 1, 1) = 0, P'(0^r, 0, 0) = 0.
\]
Thus, we can simply introduce two global dummy variables $x_{r+1}, x_{r+2}$ and include these in each constraint, and then sparsify with the predicates $P'$ instead of $P$. As a result, whichever sparsifier we obtain for $P'$ is then necessarily a sparsifier for $P$, as we can hard-code $x_{r+1} = 1, x_{r+2} = 0$, thereby recovering the behavior of the original predicate. 
\end{remark}

\begin{definition}\label{def:activeRegions}
    Let $C$ be any field-affine CSP on $n$ variables and $m$ constraints, and let $c \in C$ be a fixed constraint. Let $\pi$ be an ordering of $x_1, \dots x_n$. For a vector $x \in [0,1]^{\pi}$, we say that the active region of $x$ with respect to $c$ is 
    \[
    A(x, c) = \{\theta \in [0,1]: P(x^{(\theta)}) =1 \}.
    \]
    Note that because the active regions are defined with respect to the rounded vector $x^{(\theta)}$, the boundaries of the active region always occur at the values taken by the variables. That is, we can represent the active region as a disjoint union of intervals:
    \[
    A(x, c) = [x_{\pi_{i_1}}, x_{\pi_{i_2}}] \cup [x_{\pi_{i_3}}, x_{\pi_{i_4}}] \cup \dots [x_{\pi_{i_{2\ell -1}}}, x_{\pi_{i_{2\ell}}}].
    \]
    In particular, for a fixed ordering $\pi$, the above expression provides a general formula for the active regions of the constraint. Thus, we also refer to the active region of $c$ with respect to the permutation $\pi$ as:
        \[
    A(\pi, c) = [x_{\pi_{i_1}}, x_{\pi_{i_2}}] \cup [x_{\pi_{i_3}}, x_{\pi_{i_4}}] \cup \dots [x_{\pi_{i_{2\ell -1}}}, x_{\pi_{i_{2\ell}}}],
    \]
    where we take $x$ to be an arbitrary vector in $[0,1]^{\pi}$.

    Note again that here we are using the fact that for each constraint, $c(0^n) = c(1^n) = 0$ (as per \cref{rmk:zeroPredicate}). Otherwise, it is possible that the active region would contain an interval of the form $[0, x_{\pi_{i_1}}]$ or $[x_{\pi_{i_{\ell}}}, 1]$. But, we know that when all variables are set to $0$ or $1$, the constraint will be unsatisfied, and so the active regions will never include $\theta = 0$ or $\theta = 1$.
\end{definition}

With the notion of an active region, we can now also define crossing indices:

\begin{definition}\label{def:crossingIndexGeneral}
    Let $C$ be any field-affine CSP on $n$ variables and $m$ constraints, let $c \in C$ be a fixed constraint, and let $\pi$ be a fixed ordering of $x_1, \dots x_n$. For $i \in [n-1]$, we say that $i$ is a crossing index with respect to $c, \pi$ if the bisector $(x_{\pi_i}, x_{\pi_{i+1}})/2$ is in the active region of $c$ with respect to $\pi$.
\end{definition}

As before, we can also generalize this definition to a pair of indices $i,j \in [n-1]$:

\begin{definition}
    We say that pairs of indices $i,j \in \binom{[n-1]}{2}$ are crossing indices with respect to the constraint $c$, ordering $\pi$ if both $i, j$ individually are crossing indices with $c, \pi$.
\end{definition}

\begin{definition}\label{def:NumberCrossingGeneral}
	For a field-affine CSP $C$, a permutation $\pi$, and an index $i$, we let
    \[
	 d_{C, \pi}(i) = \sum_{c \in C} w_c \cdot \mathbf{1}[i \text{ crossing } c,\pi].
	\]
    For a pair of
    indices $i, j$, we let $d_{C, \pi}(i,j)$ be the total weight of constraints for which $i,j$ are both crossing under $\pi$. I.e.,
	\[
	 d_{C, \pi}(i,j) = \sum_{c \in C} w_c \cdot \mathbf{1}[i \text{ crossing } c,\pi \wedge j \text{ crossing } c,\pi].
	\]
\end{definition}

With this, we now introduce some auxiliary matrices:

\begin{definition}\label{def:incMatrixGeneral}
	Let $C$ be a field-affine CSP on $n$ variables, and $m$ constraints, and let $\pi$ be an ordering of the variables $x_1, \dots x_n$. Let $c$ be a single constraint, and let $x_{u_1} \leq x_{u_2} \leq \dots \leq x_{u_{|S|}}$ be the ordered variables on which $c$ operates. We say that $B^{\text{inc}}_{C, \pi} \in \R^{m \times n}$ is the \emph{incidence matrix} for $C$, where for $c \in C$, $i \in [n]$, $B^{\text{inc}}_{C, \pi}(c, i) = 0$ if $x_i$ is not present in the constraint $c$, $B^{\text{inc}}_{C, \pi}(c, i) = -1$ if $x_i$ is an upper border of part of an interval in the active region with respect to $c, \pi$, and otherwise $B^{\text{inc}}_{C, \pi}(c, i) = 1$ (when $x_i$ is a lower border of part of the active region). 
\end{definition}

\begin{claim}\label{clm:incidenceGeneral}
	For any field-affine CSP $C$ on $n$ variable and $m$ constraints, and a permutation $\pi$ on $[n]$,  for $x \in [0,1]^{\pi}$, 
    \[
    B^{\text{inc}}_{C, \pi}(x)_c = (x_{\pi_{i_1}} - x_{\pi_{i_2}}) + (x_{\pi_{i_3}} - x_{\pi_{i_4}}) + \dots + (x_{\pi_{i_{2\ell -1}}} - x_{\pi_{i_{2\ell}}}),
    \]
    for $x_{\pi_{i_1}}, \dots x_{\pi_{i_{2\ell}}}$ as defined in \cref{def:activeRegions}.
\end{claim}

\begin{proof}
    This is essentially by definition. The matrix places a $-1$ in any entry which corresponds to the upper boundary of any active region (see \cref{def:activeRegions}), and a $1$ in any lower boundary. Importantly, recall that there are no intervals of the form $[0, x_{\pi_{i_1}}]$ or $[x_{\pi_{i_{\ell}}}, 1]$ in the active regions (per \cref{rmk:zeroPredicate}).
\end{proof}

As a consequence, we also have:

\begin{claim}\label{clm:quadraticEquivalenceGeneral}
	Let $C$ be a field-affine CSP as in \cref{def:fieldaffine}, and let $W_C$ be the diagonal matrix in $\R^{m \times m}$ such that $W_C(c, c) = w_c$ (i.e., the diagonal matrix of weights of constraints). Then for $x \in [0,1]^{\pi}$
	\[
	x^T (B^{\text{inc}}_{C, \pi})^T W_C B^{\text{inc}}_{C, \pi} x = Q_C(x).
	\]
\end{claim}

\begin{proof}
	We have that 
	\[
	x^T (B^{\text{inc}}_{C, \pi})^T W_C B^{\text{inc}}_{C, \pi} x = \sum_{c \in C} w_c \cdot (B^{\text{inc}}_{C, \pi}(x)_c)^2 = \sum_{c \in C} w_c \cdot [(x_{\pi_{i_1}} - x_{\pi_{i_2}}) + (x_{\pi_{i_3}} - x_{\pi_{i_4}}) + \dots + (x_{\pi_{i_{2\ell -1}}} - x_{\pi_{i_{2\ell}}})]^2
	\]
    \[
    = \sum_{c \in C} w_c \cdot \Pr_{\theta \in [0,1]}[c(x^{(\theta)}) = 1]^2 = Q_C(x),
    \]
	where the second equality holds because of the expression for $B^{\text{inc}}_{C, \pi}(x)_c$ in \cref{clm:incidenceGeneral} and the final equality holds because of \cref{def:activeRegions}.
\end{proof}

\begin{definition}\label{def:crossMatrixfield}
    Let $C$ be a field-affine CSP as in \cref{def:fieldaffine}. We define the matrix $B^{\text{cross}}_{C, \pi} \in \R^{m \times n-1}$ such that $B^{\text{cross}}_{C, \pi}(c, i) = 1$ if $i$ is a crossing index for $c$ under $\pi$, and otherwise $B^{\text{cross}}_{C, \pi}(c, i) = 0$.
\end{definition}

Next, we will re-write $B^{\text{inc}}_{C, \pi}$ in a more convenient way:

\begin{claim}\label{clm:convenientRepfield}
	$B^{\text{inc}}_{C, \pi} = B^{\text{cross}}_{C, \pi} J P_{\pi}$, for $B^{\text{inc}}_{C, \pi}$ as defined in \cref{def:incMatrixGeneral}, $B^{\text{cross}}_{C, \pi}$ as defined in \cref{def:crossMatrixfield}, and $J, P_{\pi}$ as defined in \cref{def:specialMatrices}.
\end{claim}

\begin{proof}
	The proof is identical to the proof of \cref{clm:convenientRep}. Here we provide only some intuition for why it is true. First, it suffices to consider $\pi = \mathrm{Id}_n$, as the permutation matrices $P_{\pi}$ will invert whichever permutation is applied. Beyond this, we consider any active region as in \cref{def:activeRegions}, let us denote the active region by $[x_a, x_b]$. First, observe that when we multiply $J$ by a vector $x$, this yields a vector which looks like $[(x_2 - x_1), (x_3 - x_2), \dots, (x_n - x_{n-1})]$. Now, in an active region $[x_a, x_b]$, the crossing matrix is constructed such that entries corresponding to $(x_{a+1}-  x_a), (x_{a+2} - x_{a+1}), \dots (x_b - x_{b-1})$ are all $1$, while the other entries are $0$. In the product $B^{\text{cross}}_{C, \pi} J P_{\pi}$, we add together these consecutive terms, and the telescoping nature then leads to exactly the expression $(x_b - x_a)$.
\end{proof}

Finally, we have the following simple observation about $B^{\text{cross}}_{C, \pi}$:

\begin{claim}\label{clm:crossingNumberCrossingMatrixGeneral}
	Let $C$ be a field affine CSP as in  \cref{def:fieldaffine}. Then, for $i, j \in [n-1]$ and a permutation $\pi$, we have that $((B^{\text{cross}}_{C, \pi})^T W_C B^{\text{cross}}_{C, \pi})_{i,j} = d_{C, \pi}(i,j)$.
\end{claim}

\begin{proof}
	Observe that 
	\[
	((B^{\text{cross}}_{C, \pi})^T W_C B^{\text{cross}}_{C, \pi})_{i,j} = \sum_{c \in C} w_c \cdot \mathbf{1}[i \text{ crossing } c,\pi \wedge j \text{ crossing } c,\pi] = d_{C, \pi}(i,j),
	\]
    as per \cref{def:NumberCrossingGeneral}.
\end{proof}

\subsection{Reducing to Code Sparsification}

In this section, we show how to spectrally sparsify CSPs by using the framework of code sparsification \cite{KPS24, KPS24c}. To start, we establish the following sufficient condition for spectral sparsification:

\begin{claim}\label{clm:copositiveMatrixGeneral}
	Let $C$ be a field-affine CSP as in \cref{def:fieldaffine}. Then $\hat{C}$ (with corresponding weight matrix $W_{\hat{C}}$) is a $(1 \pm \eps)$ spectral sparsifier of $C$ if $\forall \pi$, both
	\[
	(B^{\text{cross}}_{C, \pi})^T (W_{\hat{C}} - (1 - \eps)W_C) B^{\text{cross}}_{C, \pi},
	\]
	and 
	\[
	(B^{\text{cross}}_{C, \pi})^T ((1 + \eps)W_C - W_{\hat{C}} ) B^{\text{cross}}_{C, \pi} 
	\]
	are copositive for vectors in $\R^{n-1}_+$.
\end{claim}

\begin{proof}
Indeed, let $\pi$ be a fixed permutation, and let $\hat{C}$ be created by sampling constraint in accordance with the aforementioned weighting scheme. Our goal is to show that for $x \in [0,1]^{\pi}$
\[
(1 - \eps)Q_C(x) \leq Q_{\hat{C}}(x) \leq (1 + \eps)Q_C(x).
\]
By \cref{clm:quadraticEquivalenceGeneral}, this is equivalent to showing that for $x \in [0,1]^{\pi}$
\[
(1 - \eps) x^T (B^{\text{inc}}_{C, \pi})^T W_C B^{\text{inc}}_{C, \pi} x  \leq x^T (B^{\text{inc}}_{C, \pi})^T W_{\hat{C}} B^{\text{inc}}_{C, \pi} x \leq (1 + \eps) x^T (B^{\text{inc}}_{C, \pi})^T W_C B^{\text{inc}}_{C, \pi} x,
\]
where $W_{\hat{C}} $ is now the diagonal matrix of weights corresponding to the sparsified CSP $\hat{C}$.

Now, we also invoke \cref{clm:convenientRepfield}. This implies the above copositivity is in fact equivalent to showing that 
\[
(1 - \eps) x^T P_{\pi}^{T} J^T (B^{\text{cross}}_{C, \pi})^T W_C B^{\text{cross}}_{C, \pi} J P_{\pi} x  \leq x^T P_{\pi}^{T} J^T (B^{\text{cross}}_{C, \pi})^T W_{\hat{C}} B^{\text{cross}}_{C, \pi} J P_{\pi} x 
\]
\[
\leq (1 + \eps) x^T P_{\pi}^{T} J^T (B^{\text{cross}}_{C, \pi})^T W_C B^{\text{cross}}_{C, \pi} J P_{\pi} x,
\]
for $x \in [0,1]^{\pi}$.

Now, let us focus only on one side of the above expression: namely to show that \[
(1 - \eps) x^T P_{\pi}^{T} J^T (B^{\text{cross}}_{C, \pi})^T W_C B^{\text{cross}}_{C, \pi} J P_{\pi} x  \leq x^T P_{\pi}^{T} J^T (B^{\text{cross}}_{C, \pi})^T W_{\hat{C}} B^{\text{cross}}_{C, \pi} J P_{\pi} x
\]for $x \in [0,1]^{\pi}$. This is equivalent to showing 
\[
x^T P_{\pi}^{T} J^T (B^{\text{cross}}_{C, \pi})^T (W_{\hat{C}} - (1 - \eps)W_C) B^{\text{cross}}_{C, \pi} J P_{\pi} x \geq 0.
\]

Because this is already in the form of \cref{lem:PSDtoCopos}, in order to prove the above is copositive, we must only show that 
\[
(B^{\text{cross}}_{C, \pi})^T (W_{\hat{C}} - (1 - \eps)W_C) B^{\text{cross}}_{C, \pi}
\]
is copositive for $x \in \R^{n-1}_+$, as we desire. Analogous reasoning shows the claim for the $(1 + \eps)$ side of the expression, yielding our claim. 
\end{proof}

Now, we recall the notion of a generating matrix over an arbitrary field:

\begin{definition}\label{def:genMatrixfield}
    Let $C$ be a field-affine CSP on $n$ variables with $m$ constraints. Fix a constraint $c$, such that $c(x) =1 \iff \sum_{i = 1}^n a_i x_i \neq b \mod p$, for some prime $p$. Then, the generating matrix $G$ corresponding to $C$ is in $\F_p^{m \times {n+1}}$, where for each constraint $c$, the corresponding row in $G$ places $a_i$ in the column corresponding to $x_i$, and places $-b$ in the $n+1$st column.
\end{definition}

We also use the following vector:

\begin{definition}
	Let $\pi$ be an ordering on $[n]$, and let $i \in [n]$ be an index. We say that $z_{\pi, \leq i}$ is the vector in $\zo^n$ such that $z_{\ell} = 1$ if and only if $\pi_{\ell} \leq \pi_i$.
\end{definition}

\begin{claim}\label{clm:crossingEquivalenceGeneral}
	Let $C$ be field-affine CSP on $n$ variables, $m$ constraints, and let $G$ be its canonical generating matrix. Then, for a fixed permutation $\pi$ on $n$, we have that 
	\[
	d_{C, \pi}(i) = \wt(G z_{\pi, \leq i}).
	\]
\end{claim}

\begin{proof}
	Let us consider a constraint $c \in C$. Recall that for an ordering $\pi$, the index $i$ is \say{crossing} the constraint $c$ if the sector $(x_{\pi_i}, x_{\pi_{i+1}})$ lies in one of the active regions of $c$. In particular, this means that for a choice of $\theta \in (x_{\pi_i}, x_{\pi_{i+1}})$, the constraint $c$ applied to the rounded vector $x^{(\theta)}$ evaluates to $1$. But, this rounded vector $x^{(\theta)}$ is exactly $z_{\pi, \leq i}$, as an entry in $x$ evaluates to $1$ if and only if the value at the entry is $< \theta$, which occurs iff the value at the entry is $\leq x_{\pi_i}$. Because $x \in [0,1]^{\pi}$, this occurs exactly at the indices corresponding to $z_{\pi, \leq i}$.

    So, we have established that $i$ is a crossing index for $c, \pi$ if and only if $c(z_{\pi, \leq i})$ evaluates to $1$. To conclude, we must only observe that by \cref{def:genMatrixfield}, $c(z_{\pi, \leq i}) = 1$ if and only if $(Gz_{\pi, \leq i})_c \neq 0$. The claim then follows. 
\end{proof}

As an immediate corollary, we also establish the following for pairs $i,j$:

\begin{claim}\label{clm:crossingProductGeneral}
	Let $c$ be a field-affine CSP constraint, and let $\pi$ be a permutation on $n$, let $i \neq j \in [n-1]$ be two indices, and let $G_c$ denote the row of the canonical generating matrix $G$ corresponding to constraint $c$. Then, 
	\[
	\mathbf{1}[i,j \text{ crossing } c, \pi] = \mathbf{1} [(G z_{\pi, \leq i})_c \cdot (G z_{\pi, \leq j})_c \neq 0].
	\]
\end{claim}

\begin{proof}
	\cref{clm:crossingEquivalenceGeneral} establishes that $i$ is crossing $c$ with $\pi$ if and only if $G z_{\pi, > i} \neq 0$. Thus, $i, j$ are \emph{both} crossing $c, \pi$ if and only if the \emph{product} $G z_{\pi, \geq i} \cdot G z_{\pi, \geq j} \neq 0$.
\end{proof}

So, we now define a new coding theoretic object to capture the above notion. Specifically, observe that for each constraint $c$, we are taking the \emph{and} of the evaluation of $c$ on two vectors. Because we have the relation that $c(x) = 1 \iff \sum_{i} a_i x_i \neq b \mod p$, we see that $c(x) \cdot c(y) =1 \iff (\sum_{i} a_i x_i - b) \cdot (\sum_{i} a_i y_i -b) \neq 0 \mod p$, where we use the property that $\F_p$ is a field.  

\begin{definition}\label{def:liftedGeneratingGeneral}
Let $C$ be a field-affine CSP. We let $\Phi(G)$ denote the result of lifting the generating matrix $G$ corresponding to $C$. In particular, for each constraint $c(x) = 1 \iff \sum_{i} a_i x_i \neq b \mod p$ in the original CSP, we construct the lifted quadratic equation $c(x) \cdot c(y) =1 \iff (\sum_{i} a_i x_i - b) \cdot (\sum_{i} a_i y_i -b) \neq 0 \mod p$. We re-write this right-hand side as a quadratic equation, and replace each set of variables $x_i y_j$ with a new variable $q_{i,j}$. I.e.,
\[
c(x) \cdot c(y) =1 \iff (\sum_{i} a_i x_i - b) \cdot (\sum_{i} a_i y_i -b) \neq 0 \mod p = \sum_{i,j} \alpha_{i,j} q_{i,j} - \beta \neq 0 \mod p,
\]
where we let $q\in\zo^{n^2 + 2n}$, where $q_{i,j}  = x_i, y_j$ for $i, j \in [n]$, $q_{0, j} = y_j, q_{i, 1} = x_i$.
We say the lifted generating matrix corresponding to $G$ (denoted by $\Phi(G)$) is the matrix in $\F_p^{m \times (n^2+2n + 1)}$, where the coefficient in column $i,j$ is given by $\alpha_{i,j}$, and the final column contains $- \beta$.
\end{definition}

Now, we have the following:

\begin{claim}\label{clm:preserveAndWeightsGeneral}
	Let $C$ be a field-affine CSP. Then $\hat{C}$ is a $(1 \pm \eps)$ spectral sparsifier of $C$ if $\forall \pi$, and $\forall i, j \in [n-1]$:
	\[
	\wt(Gz_{\pi, \leq i} \cdot Gz_{\pi, \leq i}) \in (1 \pm \eps) \wt(\hat{G}z_{\pi, \leq i} \cdot \hat{G}z_{\pi, \leq i}),
	\]
	where $G$ is the canonical generating matrix of $C$, and $\hat{G}$ is the canonical generating matrix of $\hat{C}$.
\end{claim}

\begin{proof}
	Recall that by \cref{clm:crossingProductGeneral}, for a permutation $\pi$, and indices $i, j$, we have that 
	\[
	\wt(Gz_{\pi, \leq i} \cdot Gz_{\pi, \leq i}) = \sum_{c \in C} w_c \cdot 
	(G z_{\pi, \leq i})_c \cdot (G z_{\pi, \leq j})_c = \sum_{c \in C} w_c \cdot \mathbf{1}[i,j \text{ crossing } c, \pi] 
	\]
	\[
	= d_{C, \pi}(i,j). 
	\]
	
	Likewise, when we use the canonical generating matrix for $\hat{C}$, we get that 
	\[
	\wt(\hat{G}z_{\pi, \leq i} \cdot \hat{G}z_{\pi, \leq i}) = d_{\hat{C}, \pi}(i,j).
	\]
	
	Finally, recall that by \cref{clm:crossingNumberCrossingMatrixGeneral}, for a fixed permutation $\pi$, we have that 
	\[
	(B^{\text{cross}}_{C, \pi})^T W_C B^{\text{cross}}_{C, \pi} = d_{C, \pi}(i,j),
	\]
	and 
	\[
	(B^{\text{cross}}_{C, \pi})^T W_{\hat{C}} B^{\text{cross}}_{C, \pi} = d_{\hat{C}, \pi}(i,j).
	\]
	
	Thus, if $\forall \pi$, and $\forall i, j \in [n-1]$
	\[
	\wt(Gz_{\pi, \leq i} \cdot Gz_{\pi, \leq  i}) \in (1 \pm \eps) \wt(\hat{G}z_{\pi, \leq  i} \cdot \hat{G}z_{\pi, \leq i}),
	\]
	this means that all entries in 
	\[(B^{\text{cross}}_{C, \pi})^T (W_{\hat{C}} - (1 - \eps)W_C) (B^{\text{cross}}_{C, \pi})^T\]
	and 
	\[
	(B^{\text{cross}}_{C, \pi})^T ((1 + \eps)W_C - W_{\hat{C}} ) B^{\text{cross}}_{C, \pi} 
	\]
	are positive, and thus the matrices are copositive for vectors in $\R^{n-1}_+$. By \cref{clm:copositiveMatrixGeneral}, this then means that $\hat{C}$ is a $(1 \pm \eps)$ spectral sparsifier of $C$.
\end{proof}

Finally, this means it suffices to find a set of weights $w_{\hat{C}}$ such that for $\forall \pi$ and $\forall i, j \in [n-1]$:
\[
\wt(Gz_{\pi, \leq i} \cdot Gz_{\pi, \leq i}) \in (1 \pm \eps) \wt(\hat{G}z_{\pi, \leq i} \cdot \hat{G}z_{\pi, \leq i}).
\]
To do this, we will once again use the lifted version of the generating matrices, and perform \say{code sparsification} on these matrices. We recall the definition of a code sparsifier from \cite{KPS24} (for the case of arbitrary fields now):

\begin{definition}
	Let $G \in \F_p^{m \times n}$ be a generating matrix. We say that $\hat{G}$ is a $(1 \pm \eps)$-code sparsifier for $G$ if the rows of $\hat{G}$ are a re-weighted subset of the rows of $G$, and $\forall x \in \F_p^n$:
	\[
	\wt(\hat{G}x) \in (1 \pm \eps) \wt(Gx).
	\]
	Here, we use $\wt$ to mean the \emph{weighted hamming weight}, in the sense that certain coordinates may be weighted more in their respective code. 
\end{definition}

The work of \cite{KPS24} showed the following:

\begin{theorem}\cite{KPS24}\label{thm:codeSparsificationGeneral}
	For any generating matrix $G \in \F_q^{m \times n}$, there is a $(1 \pm \eps)$ code-sparsifier $\hat{G}$ for $G$, such that $\hat{G} \in \F_p^{m' \times n}$, where $m' = \widetilde{O}(n \log^2(p)/ \eps^2)$.
\end{theorem}

Further, \cite{KPS24c} showed that such a code sparsifier can be computed in \emph{polynomial time} in $m, n, \eps$ (with high probability). We are now ready to provide the main proof of this section:

\begin{proof}[Proof of \cref{thm:generalCSP}]
Given the field affine CSP $C$, construct the corresponding generating matrix $G$, as well as the \emph{lifted} generating matrix $\Phi(G)$ in accordance with \cref{def:liftedGeneratingGeneral}. 
	
	Next, we invoke \cref{thm:codeSparsificationGeneral} to create a $(1 \pm \eps)$ code-sparsifier of $\Phi(G)$, which we denote by $\widehat{\Phi(G)}$. Note that because the dimension of $\Phi(G)$ is $n^2$, the number of rows (coordinates) preserved in $\widehat{\Phi(G)}$ is $\widetilde{O}(n^2  \log^2(p)/ \eps^2)$.
	
	Further, observe that every row of $\widehat{\Phi(G)}$ actually corresponds to a row of $\Phi(G)$. In particular, $\widehat{\Phi(G)} = \Phi(\widehat{G})$, where $\hat{G}$ is simply the generating matrix where each coordinate $j \in [m]$ of $G$ has the same weight as the coordinate in $\widehat{\Phi(G)}$, but where we keep row $G_j$ instead of $\Phi(G)_j$. Note that this follows because the lifting operation is applied individually to each row of $G$, and thus a collection of lifted rows is equivalent to lifting the collection of (unlifted) rows. Going forward, we will denote the CSP associated with $\hat{G}$ by $\hat{C}$.
	
	Next, for any $\pi$, and any $i, j \in [n-1]$, consider the vector $x_{\pi, i, j} \in \zo^{n^2 + 2n}$ where 
     $x_{\pi, i, j} = (z_{\pi, \leq i} \otimes z_{\pi, \leq j}) \circ z_{\pi \leq i} \circ z_{\pi \leq j}$, (where $\circ$ denotes concatenation). Observe that for any constraint $c$, (and the lifted version $\Phi(c) \in \F_2^{n^2 + 2n}$), we have
	\[
	\Phi(c) x_{\pi, i, j} = \mathbf{1}[(G z_{\pi, \leq i})_c \cdot  (G z_{\pi, \leq j})_c \neq 0] = \mathbf{1}[i,j \text{ crossing } c, \pi].
	\]
	Thus, for this vector $x_{\pi, i, j}$, we have that 
	\[
	\wt(\Phi(G) x_{\pi, i, j}) = \sum_{c \in C} w_c  \mathbf{1}[i,j \text{ crossing } c, \pi] =\wt(Gz_{\pi, > i} \cdot Gz_{\pi, > i}).
	\]
	and likewise
	\[
	\wt(\Phi(\widehat{G}) x_{\pi, i, j}) = \wt(\hat{G}z_{\pi, \leq i} \cdot \hat{G}z_{\pi, \leq i}),
	\]
	
	Because $\Phi(\hat{G})$ is a $(1 \pm \eps)$ code-sparsifier of $\Phi(\hat{G})$, and the message $x_{\pi, i, j}$ is fixed, this must mean that 
	\[
	\wt(\hat{G}z_{\pi, \leq i} \cdot \hat{G}z_{\pi, \leq i}) \in (1 \pm \eps) \wt(Gz_{\pi, \leq i} \cdot Gz_{\pi, \leq i}).
	\]
	
	Thus, we have constructed a CSP $\hat{C}$ such that $\forall \pi$, $\forall i, j \in [n-1]$, 
	\[
	\wt(Gz_{\pi, \leq i} \cdot Gz_{\pi, \leq i}) \in (1 \pm \eps) \wt(\hat{G}z_{\pi, \leq i} \cdot \hat{G}z_{\pi, \leq i}),
	\]
	and we conclude the spectral sparsification via \cref{clm:preserveAndWeightsGeneral}.
	
	The running time / probability bounds follow by using tools for efficient code sparsification from \cite{KPS24c}. 
\end{proof}

\section{Establishing Cheeger Inequalities}\label{sec:cheeger}

In this section, we show that an analog of Cheeger's inequality can hold using our spectral energy definition for CSPs. We first introduce some notation.

\subsection{Preliminaries}

Let us consider an XOR CSP where very constraint is of even arity. To start, for every constraint $c_1, \dots c_m$, where $c_i$ operates on a subset $S_i \subseteq [n]$ of the variables, we associate a corresponding \emph{hypergraph}, where the hyperedges are exactly $E = \{S_i: i \in [m]\}$. For every hyperedge $e \in E$, the weight associated to this hyperedge (denoted $w_e$) is equal to the weight assigned to the corresponding constraint that the hyperedges is from. We now introduce some further notation:

\begin{definition}
    For a vertex $v \in V$, we say that $w_V(v) = \sum_{e \in E: v \in e} w_e$. For a set $S \subseteq V$, we say that $w_V(S) = \sum_{v \in S} w_v$.

    Likewise, for any set of hyperedges $E' \subseteq E$, we say that $w_E(E') = \sum_{e \in E'} w_e$.

    Note that going forward, we will often suppress the subscript when it is clear from context whether a set is a set of vertices or a set of edges. 
\end{definition}

We also define the boundary of a set $S$:

\begin{definition}
    For a set $S \subseteq V$, $\delta S$ is the set of constraints which are satisfied by the assignment $\mathbf{1}_S$. That is to say,
    \[
    \delta S = \{e \in E: \oplus_{v \in e} (\mathbf{1}_S)_v = 1 \}.
    \]
\end{definition}

Using this notion of a boundary, we can also define \emph{expansion}:

\begin{definition}
The expansion of a set $S \subseteq V$ is defined as: 
\[
\Phi(S) = \frac{w(\delta S)}{w(S)}.
\]
\end{definition}

Finally, for the entire CSP $C$ (and its corresponding hypergraph), we define the expansion as:

\begin{definition}
    \[
    \Phi_C = \min_{S: 0 < w(S) \leq w(V) / 2} \Phi(S).
    \]
\end{definition}

Next, we can also define the discrepancy ratio, which will be important for our version of the Cheeger inequality.  

\begin{definition}
    For a vector $f \in \R^V$ (i.e., an assignment of potentials to the vertices), we say that 
    \[
    D_w(f) = \frac{\sum_{c \in C} w_c \cdot Q_c(f)^2}{\sum_{u \in V} w_u f_u},
    \]
    where $Q_c(f)$ denotes the energy function of a single constraint $c$. I.e., if $f \in [0,1]^n$, then $Q_c(f) = \Pr_{\theta}[c(f^{(\theta)}) = 1] $, and otherwise, if $f \in R^n$, we set $Q_c(f) = (\max(f) - \min(f)) \cdot \Pr_{\theta}[c(\hat{f}^{(\theta)}) = 1]$, where $\hat{f} = \frac{f - \mathbf{1} \cdot \min(f)}{\max(f) - \min(f)}$ is simply the scaled down and translated version of $f$ such that it fits in the interval $[0,1]$.
\end{definition}

We also require a notion of the \emph{weighted space}. For $f, g \in \R^V$, we say that $\langle f, g \rangle_w = f^T W G$, where $W \in \R^{n \times n}$ is the diagonal matrix with $W_{u,u} = w_u$. 

\begin{definition}
    The \emph{normalized} discrepancy ratio of a vector $x \in \R^V$ is given by 
    \[
    \mathcal{D}(x) = D_w(W^{-1/2} x).
    \]
\end{definition}

\begin{definition}\label{def:eigenvalue}
    We let 
    \[
    \gamma_2 = \min_{0 \neq x \perp W^{1/2} \mathbf{1}} \mathcal{D}(x)
    \]
    denote the second eigenvalue of the normalized discrepancy ratio. 
\end{definition}

\subsection{Groundwork}

Now, we will begin proving some auxiliary lemmas towards our Cheeger inequality. Ultimately, what we will show is the following:

\begin{theorem}
    Given an XOR CSP $C$ where each constraint is of even size and maximum arity $\ell$,
    \[
    \frac{\gamma_2}{2} \leq \Phi_C \leq \left (2 \sqrt{\ell/2} + 1 \right) \sqrt{\gamma_2},
    \]
    where $\gamma_2$ is the eigenvalue of the XOR-CSP Laplacian defined in \cref{def:eigenvalue}. 
\end{theorem}

\begin{remark}
    Note that the reason we only use constraints of even arity is that otherwise we cannot translate an assignment by a multiple of the all $1$'s vector without loss of generality. This is because any XOR of even arity always evaluates to $0$ on the all $1$'s vector, while any vector of odd arity evaluates to $1$.
\end{remark}

As a first, step, let us consider the following claim we wish to prove: 

\begin{claim}\label{clm:rounding}
    Let $C$ be an XOR CSP where each constraint is of even size, and maximum arity $\ell$, and let $f \in [0,1]^V$ be non-zero. Then, there exists a set $S \subseteq \Supp(f)$ such that 
    \[
    \Phi(S) \leq \frac{\sum_{c \in C} w_c \cdot \Pr_{\theta \sim [0,1]}[c(f^{(\theta)}) = 1]}{\sum_{u \in V} w_u f_u}.
    \]
\end{claim}

\begin{proof}
    Let us assume for ease of notation that the entries of $f$ are sorted. I.e., that $f_1 \leq f_2 \leq \dots \leq f_n$. Observe then that we can re-write the numerator as
    \[
    \sum_{c \in C} w_c \cdot \Pr_{\theta \sim [0,1]}[c(f^{(\theta)}) = 1] = \sum_{i = 1}^n (f_i - f_{i-1}) \cdot (\sum_{c \in C} w_c \cdot \Pr_{\theta \sim [f_{i-1},f_i]}[c(f^{(\theta)}) = 1]) 
    \]
    \[
    = \sum_{i = 1}^n \left [\left(f_i - f_{i-1}\right)\left(\sum_{c \in C} w_c \cdot c(e^{\geq i}) \right)\right],
    \]
    where we are using $e_{\geq i}$ to denote the vector in $\R^n$ which is $1$ in coordinates $\geq i$, and $0$ otherwise. 

    Likewise, we can write the denominator as 
    \[
    \sum_{u \in V } w_u f_u = \sum_{i = 1}^n \left [ \left(f_{i} - f_{i-1}\right) \cdot \left(\sum_{u \in V: u \geq i} w_u\right) \right].
    \]

    Putting this together, this means that 
    \[
    \frac{\sum_{c \in C} w_c \cdot \Pr_{\theta \sim [0,1]}[c(f^{(\theta)}) = 1]}{\sum_{u \in V} w_u f_u} = \frac{\sum_{i = 1}^n \left [\left(f_i - f_{i-1}\right)\left(\sum_{c \in C} w_c \cdot c(e^{\geq i}) \right)\right]}{\sum_{i = 1}^n \left [ \left(f_{i} - f_{i-1}\right) \cdot \left(\sum_{u \in V: u \geq i} w_u\right) \right]}.
    \]

    In particular, there must be an index $i$ such that
    \[
    \frac{\left [\left(f_i - f_{i-1}\right)\left(\sum_{c \in C} w_c \cdot c(e^{\geq i}) \right)\right]}{\left [ \left(f_{i} - f_{i-1}\right) \cdot \left(\sum_{u \in V: u \geq i} w_u\right) \right]} \leq  \frac{\sum_{c \in C} w_c \cdot \Pr_{\theta \sim [0,1]}[c(f^{(\theta)}) = 1]}{\sum_{u \in V} w_u f_u}.
    \]

    This means that 
    \[
    \frac{\left(\sum_{c \in C} w_c \cdot c(e^{\geq i}) \right)}{  \left(\sum_{u \in V: u \geq i} w_u\right) } \leq  \frac{\sum_{c \in C} w_c \cdot \Pr_{\theta \sim [0,1]}[c(f^{(\theta)}) = 1]}{\sum_{u \in V} w_u f_u}.
    \]

    However, this expression on the left is exactly $\Phi(S)$, where $S = \{ u \in V: u \geq i\}$. Thus, for this set $S$ we get exactly the desired claim. Observe that for general orderings, instead of peeling off the indices $i$ in the order $1, 2, \dots n$, we would instead peel them off based on the ordering $f$ satisfies. 
\end{proof}

Next, we need to show the following claim: 

\begin{claim}\label{clm:upperBound}
    Let $C$ be an XOR CSP where each constraint is of even size, and maximum arity $\ell$, and let $f \in \R^V$ be a non-zero assignment to the vertices such that $f \perp_w \mathbf{1}$. Then, there exists a set $S$ such that $w(S) \leq w(V) / 2$, and 
    \[
    \Phi(S) \leq D_w(f) + 2 \sqrt{D_w(f)}.
    \]
\end{claim}

\begin{proof}
    We follow the path of \cite{CLTZ18}. First, let $g = f + c \cdot \mathbf{1}$ be such that both $w(\Supp(g^+))$ and $w(\Supp(g^-))$ are at most $w(V) /2$.

    Because we are told that $f \perp_w \mathbf{1}$, this means that 
    \[
    \langle g, \mathbf{1} \rangle_w =  c \langle \mathbf{1}, \mathbf{1} \rangle_w .
    \]
    Thus, we get that 
    \[
    \langle f, f \rangle_w = \langle g, g \rangle_w - 2c \langle g, \mathbf{1} \rangle_2 + c^2 \langle \mathbf{1}, \mathbf{1} \rangle_w =  \langle g, g \rangle_w - c^2 \langle \mathbf{1}, \mathbf{1} \rangle_w \leq \langle g, g \rangle_w.
    \]

    This means that 
    \[
    D_w(f) = \frac{\sum_{c \in C} w_c \cdot Q_c(g)^2}{\sum_{u \in V} w_u f_u^2} \geq \frac{\sum_{c \in C} w_c \cdot Q_c(g)^2}{\sum_{u \in V} w_u g_u^2} = D_w(g).
    \]

    Next, we want to decompose the above expression into $g^+$ and $g^-$. Specifically, observe that 
    \[
    Q_c(g)^2 \geq Q_c(g^{+})^2+ Q_c(g^{-})^2,
    \]
    as $Q_c(g) =Q_c(g^{+}) + Q_c(g^{-})$, and all are non-negative.

    Thus, 
    \[
    D_w(f) \geq D_w(g) = \frac{\sum_{c \in C} w_c \cdot Q_c(g)^2}{\sum_{u \in V} w_u g_u^2}
    \]
    \[
    \geq \frac{\sum_{c \in C} w_c \cdot Q_c(g^+)^2 + \sum_{c \in C} w_c \cdot Q_c(g^-)^2}{\sum_{u \in V} w_u (g_u^+)^2 + \sum_{u \in V} w_u (g_u^-)^2}
    \]
    \[
    \geq \min(D_w(g^+), D_w(g^-))
    \]

    Let $h$ be the vector corresponding to the minimizer of $\{g^+, g^-\}$. By \cref{clm:simplificationProof}, we then have that
    \[
    \sum_{c \in C} w_c \cdot Q_c(h^2) \leq \sum_{c \in C}  w_c \cdot Q_c(h)^2 + 2 \sqrt{\ell/2} \cdot \sqrt{\sum_c w_c Q_c(h)^2} \cdot \sqrt{\sum_{u \in V} w_u \cdot h_u^2}.
    \]

Dividing by $\sum_{u \in V} w_u h_u^2$, and using that $D_w(h) \leq D_w(f)$, we get that
\[
\frac{\sum_{c \in C} w_c Q_c(h^2)}{\sum_{u \in V} w_u h_u^2} \leq D_w(h) + 2\sqrt{\ell/2} \sqrt{D_w(h)} \leq D_w(f) + 2\sqrt{\ell/2} \sqrt{D_w(f)}.
\]

Invoking \cref{clm:rounding} with the vector $h^2$, we get that there exists a vector $S \subseteq \Supp(h)$ such that 
\[
\Phi(S) \leq D_w(f) + 2\sqrt{\ell/2} \sqrt{D_w(f)},
\]
and $w(S) \leq \frac{w(V)}{2}$.
\end{proof}

Finally, we can conclude:

\begin{theorem}
    Given an XOR CSP $C$ where each constraint is of even size and maximum arity $\ell$,
    \[
    \frac{\gamma_2}{2} \leq \Phi_C \leq \left (2 \sqrt{\ell/2} + 1 \right) \sqrt{\gamma_2},
    \]
    where $\gamma_2$ is the eigenvalue of the XOR-CSP Laplacian defined in \cref{def:eigenvalue}. 
\end{theorem}

\begin{proof}
    First, let $S$ be any set such that $w(S) \leq w(V) / 2$. Let $g$ be the indicator vector for $S$. Let $f$ be the component of $g$ orthogonal to $\mathbf{1}$. Then, $g = f + c \cdot 1$, where $c = w(S) / w(V)$.

    One can check that $\langle f, f \rangle_w \geq w(S) / 2$ (see \cite{CLTZ18}). We then get that 
    \[
    \gamma_2 \leq D_w(f) = \frac{\sum_{c \in C} w_c \cdot Q_c(g) }{\langle f, f \rangle_w}
    \]
    \[
    \leq \frac{2w(\delta S)}{w(S)} = 2 \Phi(S),
    \]
    where we use that $g$ is an indicator vector of the set $S$, and hence $Q_c(g) = \mathbf{1}[c(\mathbf{1}_S) = 1]$.
    Because this is true for any set $S$ which satisfies $w(S) \leq w(V) / 2$, we get that $\Phi_C \geq \gamma_2 / 2$. On the other hand, via \cref{clm:upperBound}, we get that 
    \[
    \Phi_C \leq D_w(f) + 2 \sqrt{\ell/2}\sqrt{D_w(f)} \leq (2 \sqrt{\ell/2} + 1)\sqrt{\gamma_2}.
    \]
    Together, these yield our desired claim. 
\end{proof}

\bibliographystyle{alpha}
\bibliography{ref}

\appendix

\section{Elaborating on the Proof of \cref{clm:upperBound}}

Recall that we are dealing with a vector $h$ here which is the minimizer of $\{g^+, g^-\}$ as in the proof of \cref{clm:upperBound}, and we are analyzing the expression $\sum_{c \in C} w_c Q_c(h^2)$, where each constraint $c$ is an XOR of even size. We claim that we have the following simplification:

\begin{claim}\label{clm:simplificationProof}
    Let $C$ be an XOR instance on $n$ variables, where each constraint has even arity, and maximum size $\ell$, and let $h \in \R_{\geq 0}^n$ be an assignment to these $n$ variables. Then, 
    \[
    \sum_{c \in C} w_c \cdot Q_c(h^2) \leq \sum_{c \in C}  w_c \cdot Q_c(h)^2 + 2 \sqrt{\ell/2} \cdot \sqrt{\sum_c w_c Q_c(h)^2} \cdot \sqrt{\sum_{u \in V} w_u \cdot h_u^2}.
    \]
\end{claim}

\begin{proof}
    First, let us assume without loss of generality that the variables are in sorted order, that is to say, that $h_1 \leq h_2 \leq \dots \leq h_n$. Note that this is without loss of generality because our proof can be expressed in terms of the ordering of the variables, but we simply focus on this case for ease of notation. Now, our first claim is that for a single constraint $c$ of size $\ell$:
    \[
Q_c(h^2) \leq Q_c(h)^2 + 2 \cdot Q_c(h) \cdot \left ( \sum_{i = 1, 3, \dots \ell -1} \min^{(i)}_{u \in c} h_u \right ).
    \]
    Here we use $\min^{(i)}$ to denote the $i$th smallest element in a collection. To simplify notation however, we will assume that the constraint $c$ is instead operating on the variables $h_1, \dots h_{\ell}$. In this way, the $i$th smallest element is exactly $h_i$. Likewise, we can also simplify the expression $Q_c(h)$, as this will now be exactly 
    \[
    Q_c(h) = h_{\ell} - h_{\ell-1} + \dots + h_2 - h_1 = \sum_{i = 1}^{\ell} (-1)^i h_i.
    \]
    Thus, our goal is exactly to show that 
    \[
    Q_c(h^2) = \sum_{i = 1}^{\ell} (-1)^i h_i^2 \]
    \[
    \leq \left ( \sum_{i = 1}^{\ell} (-1)^i h_i \right )^2 + 2 \cdot \left ( \sum_{i = 1}^{\ell} (-1)^i h_i \right ) \cdot \left ( \sum_{i = 1, 3, \dots \ell-1} h_i \right ) =  Q_c(h)^2 + 2 \cdot Q_c(h) \cdot \left ( \sum_{i = 1, 3, \dots \ell -1} \min^{(i)}_{u \in c} h_u \right ) .
    \]
    So, now we must only match terms, but there are several cases to consider:
    \begin{enumerate}
        \item Let us analyze terms of the form $h_ah_b$ where $a$ is even and $b$ is odd. No such term is present on the left hand side of the above expression. On the right hand side, the expression $\left( \sum_{i = 1}^{\ell} (-1)^i h_i \right )^2$ contributes a $-2$ coefficient to any such term, while $2 \cdot \left ( \sum_{i = 1}^{\ell} (-1)^i h_i \right ) \cdot \left ( \sum_{i = 1, 3, \dots \ell-1} h_i \right )$ contributes a $+2$ to any such term, and so the total coefficient is $0$. 
        \item Now, let us consider terms of the form  $h_ah_b$ where $a \neq b$, but $a$ and $b$ are both even. The left hand side again does not contribute any such terms. On the right hand side, the expression $\left( \sum_{i = 1}^{\ell} (-1)^i h_i \right )^2$ contributes a $+2$ coefficient to any such term, while $2 \cdot \left ( \sum_{i = 1}^{\ell} (-1)^i h_i \right ) \cdot \left ( \sum_{i = 1, 3, \dots \ell-1} h_i \right )$ contributes $0$. Hence, the final coefficient is $+2$.
        \item Now, let us consider terms of the form  $h_ah_b$ where $a \neq b$, but $a$ and $b$ are both odd. The left hand side again does not contribute any such terms. On the right hand side, the expression $\left( \sum_{i = 1}^{\ell} (-1)^i h_i \right )^2$ contributes a $+2$ coefficient to any such term, while $2 \cdot \left ( \sum_{i = 1}^{\ell} (-1)^i h_i \right ) \cdot \left ( \sum_{i = 1, 3, \dots \ell-1} h_i \right )$ contributes $-2$. Hence, the final coefficient is $0$.
        \item Next, let us consider the terms $x_a^2$, where $a$ is odd. The left hand side contributes a coefficient of $-1$ in such a case. On the right hand side $\left( \sum_{i = 1}^{\ell} (-1)^i h_i \right )^2$ contributes a coefficient of $+1$, while $2 \cdot \left ( \sum_{i = 1}^{\ell} (-1)^i h_i \right ) \cdot \left ( \sum_{i = 1, 3, \dots \ell-1} h_i \right )$ contributes $-2$, leading to a final coefficient of $-1$. 
        \item Finally, let us consider the terms $x_a^2$, where $a$ is even. The left hand side contributes a coefficient of $1$ in such a case. On the right hand side $\left( \sum_{i = 1}^{\ell} (-1)^i h_i \right )^2$ contributes a coefficient of $+1$, while $2 \cdot \left ( \sum_{i = 1}^{\ell} (-1)^i h_i \right ) \cdot \left ( \sum_{i = 1, 3, \dots \ell-1} h_i \right )$ contributes $0$, leading to a final coefficient of $1$. 
    \end{enumerate}

    So, we see that 
    \[
    \left ( \sum_{i = 1}^{\ell} (-1)^i h_i \right )^2 + 2 \cdot \left ( \sum_{i = 1}^{\ell} (-1)^i h_i \right ) \cdot \left ( \sum_{i = 1, 3, \dots \ell-1} h_i \right ) - \sum_{i = 1}^{\ell} (-1)^i h_i^2
    \]
    \[
    = \sum_{a \neq b, \text{even}} h_a h_b \geq 0,
    \]
    as we assumed every entry has the same sign. So, we have established that 
    \[
    Q_c(h) \leq Q_c(h)^2 + 2 \cdot Q_c(h) \cdot \left ( \sum_{i = 1, 3, \dots \ell -1} \min^{(i)}_{u \in c} h_u \right ) .
    \]
    Next, by taking a sum over the constraints, we get that 
    \[
    \sum_{c \in C} w_c \cdot Q_c(h^2) \leq \sum_{c \in C} w_c \cdot\left ( Q_c(h)^2 + 2 \cdot Q_c(h) \cdot \left ( \sum_{i = 1, 3, \dots \ell -1} \min^{(i)}_{u \in c} h_u \right )   \right ).
    \]
    Now, by Cauchy-Schwartz, we get that 
    \[
    \sum_{c \in C} w_c \cdot \left ( Q_c(h)^2 + 2 \cdot Q_c(h) \cdot \left ( \sum_{i = 1, 3, \dots \ell -1} \min^{(i)}_{u \in c} h_u \right )   \right )
    \]
    \[
    \leq  \sum_{c \in C} w_c \cdot Q_c(h)^2 + 2 \cdot \sqrt{\sum_c w_c Q_c(h)^2} \cdot \sqrt{\sum_{c \in C} w_c \left ( \sum_{i = 1, 3, \dots \ell -1} \min^{(i)}_{u \in c} h_u \right )^2}.
    \]
    Next, we claim that 
    \[
    \left ( \sum_{i = 1, 3, \dots \ell -1}\min^{(i)}_{u \in c} h_u \right )^2 \leq \ell/2 \cdot \sum_{i \in c}^{\ell} h_i^2. 
    \]
    Again, let us assume that $c$ is a constraint of arity $\ell$, and the vector $h$ is sorted: $h_1 \leq h_2 \leq \dots \leq h_{\ell}$. Then, the left hand side is exactly 
    \[
    (h_1 + h_3 + h_5 + \dots + h_{\ell-1})^2,
    \]
    and the right hand side is $\ell/2 \cdot \sum_i h_i^2$. Then, we have the simple fact that $\ell/2 \cdot (h_{\ell}^2 + h_{\ell-1}^2) \geq \ell h_{\ell-1}^2 \geq 2 \cdot \sum_{i = 1, 3, 5 \dots \ell-1} h_{\ell-1}^2 \geq 2 \cdot h_{\ell-1} \cdot \sum_{i = 1, 3, 5 \dots \ell-1} h_{i}$. Likewise, we can repeat this argument for 
    \[
    \ell/2 \cdot (h_{\ell-2}^2 + h_{\ell-3}^2) \geq 2 \cdot h_{\ell-3} \cdot \sum_{i = 1, 3, 5 \dots \ell-3} h_{i},
    \]
    and so on. This establishes that $\left ( \sum_{i = 1, 3, \dots \ell -1}\min^{(i)}_{u \in c} h_u \right )^2 \leq \ell/2 \cdot \sum_{i \in c}^{\ell} h_i^2$.

    Thus, we see that 
    \[
    \sum_{c \in C} w_c \cdot  Q_c(h)^2 + 2 \cdot \sqrt{\sum_c w_c Q_c(h)^2} \cdot \sqrt{\sum_{c \in C} w_c \left ( \sum_{i = 1, 3, \dots \ell -1} \min^{(i)}_{u \in c} h_u \right )^2} 
    \]
    \[
    \leq  \sum_{c \in C}  w_c \cdot Q_c(h)^2 + 2 \cdot \sqrt{\sum_c w_c Q_c(h)^2} \cdot \sqrt{\sum_{c \in C} w_c\cdot  (\ell/2) \cdot \sum_{i \in c}^{\ell} h_i^2}
    \]
    \[
    = \sum_{c \in C} w_c \cdot Q_c(h)^2 + 2 \sqrt{\ell/2} \cdot \sqrt{\sum_c w_c Q_c(h)^2} \cdot \sqrt{\sum_{c \in C} w_c \cdot \sum_{i \in c}^{\ell} h_i^2}
    \]
    \[
    = \sum_{c \in C} w_c \cdot  Q_c(h)^2 + 2 \sqrt{\ell/2} \cdot \sqrt{\sum_c w_c Q_c(h)^2} \cdot \sqrt{\sum_{u \in V} w_u \cdot h_u^2}.
    \]
    This concludes the claim. 
\end{proof}

\end{document}